\documentclass[manuscript]{acmart}

\usepackage{algorithmic}
\usepackage{subcaption}
\usepackage{multirow}
\usepackage{hhline}
\usepackage{csquotes}
\usepackage{enumitem}
\usepackage{array}
\usepackage{footmisc}
\usepackage[table]{xcolor}
\usepackage{colortbl}
\usepackage{graphbox}
\usepackage{pifont}

\definecolor{pilot}{HTML}{a9cff5}   % azzurro vivo
\definecolor{copilot}{HTML}{ffcd82} % arancione brillante
\definecolor{overlap}{HTML}{bcf7bc}  % verde chiaro

\definecolor{green}{RGB}{5,150,105}
\definecolor{red}{RGB}{220,38,38}
\definecolor{orange}{RGB}{234,88,12}
\definecolor{sectiongray}{gray}{0.9}
\definecolor{lightrowgray}{HTML}{F2F2F2}

\newcommand{\system}{\textit{GamePals}}
\newcommand{\systemFullName}{\textit{GamePals} framework}
\newcommand{\systemFullNameTitle}{\textit{GamePals} Framework}

\newcommand{\hc}{\paragraph{\textcolor{red}{Human cooperation}}}
\newcommand{\pa}{\paragraph{\textcolor{blue}{Partial automation}}}

\newcommand{\quots}[1]{``#1''}
\newcommand{\cut}[1]{}

\newcommand{\ParticipantTwo}{\textbf{P1}}
\newcommand{\ParticipantThree}{\textbf{P2}}
\newcommand{\ParticipantFour}{\textbf{P3}}
\newcommand{\ParticipantFive}{\textbf{P4}}
\newcommand{\ParticipantSix}{\textbf{P5}}
\newcommand{\ParticipantSeven}{\textbf{P6}}
\newcommand{\ParticipantEight}{\textbf{P7}}
\newcommand{\ParticipantNine}{\textbf{P8}}
\newcommand{\ParticipantTen}{\textbf{P9}}
\newcommand{\ParticipantEleven}{\textbf{P10}}
\newcommand{\ParticipantTwelve}{\textbf{P11}}

\newcommand{\ParticipantFifteen}{\textbf{P12}}
\newcommand{\ParticipantSixteen}{\textbf{P13}}

\newcommand{\checkgreen}{\footnotesize \textcolor{green}{\ding{51}}}
\newcommand{\xred}{\textcolor{red}{\ding{55}}}
\newcommand{\revisionCHI}[1]{#1} % Uncoment and comment below to remove highlighting

\begin{document}

\title{Video Game Accessibility through Shared Control for People with Upper-Limb Impairments}
%\title{Comparing human cooperation and partial automation in game accessibility through shared control for people with upper-limb impairments}
%\title{Game accessibility for people with upper-limb impairments through shared control: comparing human cooperation and partial automation}

\author{Dragan Ahmetovic}
\email{dragan.ahmetovic@unimi.it}
\orcid{0000-0001-5745-1230}
\affiliation{%
  \institution{Università degli Studi di Milano, Department of Computer Science}
  \city{Milan}
  \country{Italy}
}

\author{Matteo Manzoni}
\email{matteo.manzoni@unimi.it}
\orcid{0009-0002-1946-5919}
\affiliation{%
  \institution{Università degli Studi di Milano, Department of Computer Science}
  \city{Milan}
  \country{Italy}
}

\author{Filippo Corti}
\email{filippo.corti1@studenti.unimi.it}
\orcid{0009-0001-2493-6132}
\affiliation{%
  \institution{Università degli Studi di Milano, Department of Computer Science}
  \city{Milan}
  \country{Italy}
}

\author{Sergio Mascetti}
\email{sergio.mascetti@unimi.it}
\orcid{0000-0002-8416-4023}
\affiliation{%
  \institution{Università degli Studi di Milano, Department of Computer Science}
  \city{Milan}
  \country{Italy}
}

\renewcommand{\shortauthors}{Ahmetovic et al.}

\begin{CCSXML}
<ccs2012>
   <concept>
       <concept_id>10010405.10010476.10011187.10011190</concept_id>
       <concept_desc>Applied computing~Computer games</concept_desc>
       <concept_significance>300</concept_significance>
       </concept>
   <concept>
       <concept_id>10003120.10011738.10011775</concept_id>
       <concept_desc>Human-centered computing~Accessibility technologies</concept_desc>
       <concept_significance>500</concept_significance>
       </concept>
   <concept>
       <concept_id>10003120.10011738.10011773</concept_id>
       <concept_desc>Human-centered computing~Empirical studies in accessibility</concept_desc>
       <concept_significance>500</concept_significance>
       </concept>
   <concept>
       <concept_id>10003120.10003121.10003124.10011751</concept_id>
       <concept_desc>Human-centered computing~Collaborative interaction</concept_desc>
       <concept_significance>500</concept_significance>
       </concept>
   <concept>
       <concept_id>10003456.10010927.10003616</concept_id>
       <concept_desc>Social and professional topics~People with disabilities</concept_desc>
       <concept_significance>300</concept_significance>
       </concept>
 </ccs2012>
\end{CCSXML}

\ccsdesc[300]{Applied computing~Computer games}
\ccsdesc[500]{Human-centered computing~Accessibility technologies}
\ccsdesc[500]{Human-centered computing~Empirical studies in accessibility}
\ccsdesc[500]{Human-centered computing~Collaborative interaction}
\ccsdesc[300]{Social and professional topics~People with disabilities}

\keywords{Shared control; partial automation; human cooperation; accessible gaming.}

\begin{abstract} %dopo check con chatgpt
Interacting with video games is challenging for people with upper-limb impairments, especially when multiple hand-based inputs are required in rapid succession. Human cooperation, where another person assists the player, has been proposed as a solution, but it is limited by copilot availability and co-location.
An alternative is partial automation, where the player is assisted by a software agent.
%In this work, w
We present a study with 13 participants with upper-limb impairments, investigating how they collaborate with a copilot in both human cooperation and partial automation. The experiment is supported by GamePals, a configurable framework we developed to enable both human cooperation and partial automation in existing third-party video games.
\end{abstract}

\maketitle

\section{Introduction}
\label{sec:introduction}
Most video games are played using hands~\cite{martinez2024playing, yuan2011game}, posing accessibility barriers for people with upper-limb impairments~\cite{bierre2005game, dalgleish2023can, brown2021designing, ahmetovic2026shared}.
Solutions such as input remapping~\cite{joyToKey,reWASD}, adaptive controllers~\cite{xboxAdaptiveController,ps5accessController,flexController}, or alternative interaction modalities~\cite{ahmetovic2021replay,manzoni2024personalized,voiceAttack,playAbility}, can mitigate some accessibility problems but are not always sufficient~\cite{cimolino2021role}.
This is particularly evident in video games requiring rapid execution of multiple actions~\cite{brown2021designing,ahmetovic2026shared}.
In such cases, shared control is an approach that %has emerged as a promising assistive technology that 
can be used to reduce the number of commands a player must manage by delegating some of them to a copilot~\cite{dalgleish2023can,ahmetovic2026shared}.

One type of shared control is \textit{human cooperation}, where the player is assisted by a human copilot.
%This modality is already available on some commercial gaming platforms~\cite{xboxControllerAssist, ps5assistController, titanTwo} and can be used as an assistive technology for all the games available on them.
This modality is already available on some commercial gaming platforms such as Xbox~\cite{xboxControllerAssist} and PlayStation~\cite{ps5assistController}.
%, and can be used for the games available on them.
However, it requires the presence of another person, thus introducing dependency on their availability and skills, ultimately reducing players' autonomy~\cite{ahmetovic2026shared}.
\textit{Partial automation} is another type of shared control, in which the player is assisted by a software copilot, removing the dependency on others~\cite{ahmetovic2026shared}.
However, partial automation is currently limited to a few commercial games~\cite{callOfDuty, marioKart, bayonetta, bayonetta2} and research prototypes~\cite{cimolino2021role, medeiros2015developing, hwang2017game, cimolino2023automation, cechanowicz2014improving}, while a generalizable solution
%to provide accessibility support for video games through partial automation
remains unavailable.

%One research gap in this domain is, therefore, how to design a generalizable assistive technology for partial automation.
%The design of such a system can be informed by analyzing how existing human cooperation and partial automation solutions are used in practice.
%However, although human cooperation and partial automation have been studied individually as assistive technologies~\cite{gonccalves2021exploring, cimolino2021role, medeiros2015developing, hwang2017game}, a second research gap is that no studies have examined how the support provided by human cooperation and partial automation differs.

To inform the design of a generalizable assistive technology for partial automation, we explore how human cooperation and partial automation solutions can be used in practice.
%Indeed, while specific assistive technologies based on human cooperation and partial automation have been studied individually~\cite{gonccalves2021exploring, cimolino2021role, medeiros2015developing, hwang2017game}, no prior studies have examined how shared control support is used by end users and how the two modalities differ in providing this support.
%To address these gaps, 
Specifically, we pose the following research questions:
\begin{itemize}
    \item[\textbf{RQ1}:] How do players perceive gaming in human cooperation and partial automation?
    \item[\textbf{RQ2}:] How do players collaborate in human cooperation and partial automation, and what are the differences?
    \item[\textbf{RQ3}:] What functionalities should partial automation provide to assist players with upper-limb impairments?
\end{itemize}

To answer these research questions, as a first contribution, we conduct a user study involving $13$ participants with upper-limb impairments, investigating how gaming unfolds in human cooperation and partial automation and how players collaborate with the copilot in these two modalities.
Specifically, we focus on understanding how the (human or software) copilot assists the pilot with the game controls, in order to provide \textit{assistance by playing}~\cite{ahmetovic2026shared}.
%
%To support the study, as a second contribution, we developed \system{}, a novel, configurable open-source research probe\footnote{\label{fn:source-code}\url{https://anonymous.4open.science/r/GamePals-B88B}} that enables both human cooperation and partial automation in existing commercial video games.
%Drawing on the features of state-of-the-art shared control systems, \system{} implements them into a single research probe that allows researchers to experiment with a wide variety of shared control configurations.
%Indeed, thanks to \system{}, we were able to conduct the study using the popular commercial video game \textit{Rocket League}~\cite{rocketLeague}.

To support the study, as a second contribution, we developed \system{}, a configurable open-source research probe\footnote{\label{fn:source-code}\url{https://anonymous.4open.science/r/GamePals-B88B}} that enables both human cooperation and partial automation in existing commercial video games.
\system{} consolidates key features of prior shared-control systems into a single framework, allowing us to conduct the study with the commercial video game \textit{Rocket League}~\cite{rocketLeague}.

Results show that both shared control modalities enable participants to play by reducing the number of controls assigned to them.
The analysis also uncovers differences between the two modalities: human cooperation promotes sociality but also dependency; partial automation, instead, increases autonomy
%by removing the need for human support 
and allows some players to feel more competent, though it sometimes also generates \textit{collaboration confusion} when players misattribute actions.

After presenting \system{} and discussing the main findings, we outline key design recommendations for future partial automation systems, including interpretability of the pilot’s actions and adaptability to individual abilities.

\section{Related Work}
\label{sec:related}

We analyze existing video game assistive technologies for people with upper-limb impairments (Section~\ref{ssec:related-accessibility}) and present prior shared control solutions for video games (Section~\ref{ssec:shared-control}).

\subsection{Video Game Accessibility for People with Upper-Limb Impairments}
\label{ssec:related-accessibility}

For people with upper-limb impairments, accessing video games is challenging~\cite{martinez2024playing,bierre2005game,dalgleish2023can,brown2021designing,ahmetovic2026shared,yuan2011game}.
\revisionCHI{The severity of this challenge depends on the player's specific abilities and the game being played~\cite{yuan2011game}.
In particular, one major factor contributing to the scarce accessibility of video games is that modern real-time games like first-person shooters require strict reaction-time and simultaneous inputs~\cite{martinez2024playing,yuan2011game}.
As a result, modern gamepads~\cite{xboxController,  ps5Controller}
%, such as the \textit{Xbox}~\cite{xboxController} and \textit{PlayStation}~\cite{ps4Controller, ps5Controller} controllers,
are designed for players to coordinate multiple inputs simultaneously~\cite{brown2013evaluating}.
Specifically, users are expected to use both hands to control two analog sticks with their thumbs, press up to eight face buttons, and operate four back buttons with their index and middle fingers.
This layout implicitly assumes users to have similar hand size and morphology~\cite{brown2013evaluating}, to be able to perform coordinated input across both hands~\cite{dalgleish2023can}, and execute rapid or precisely timed movements~\cite{dalgleish2023can}.}

To overcome these accessibility challenges, players with upper-limb impairments adopt a wide range of strategies.
One possibility is customizing controls through input remapping~\cite{dalgleish2023can,brown2021designing}, using in-game options or external tools, when such options are unavailable~\cite{joyToKey,reWASD}.
%for example by remapping the most important commands to the left side of the controller for players unable to use their right hand~\cite{ahmetovic2026shared}.
For example, a player who cannot use their right hand may remap the most important commands to the left side of the controller~\cite{ahmetovic2026shared}.
%This process can happen through in-game options or with the help of external tools, like operating system services or third-party software~\cite{joyToKey,reWASD}.
\revisionCHI{However, input remapping still requires players to physically operate the controller to control all game actions.
When a player’s impairment prevents them from reliably activating these inputs, remapping becomes insufficient.}
Another technique is using alternative interaction modalities as input via hardware and software solutions.
For example, prior works have proposed voice commands~\cite{voiceAttack,ahmetovic2021replay} and facial expressions~\cite{manzoni2024personalized,playAbility} as alternative interaction modalities.
\revisionCHI{While these modalities can improve game accessibility in some cases, they can be slower than traditional controller inputs~\cite{ahmetovic2021replay}.
Additionally, the number of controls players must manage remains unchanged; thus, mapping a distinct alternative action to every required input can be challenging.
%This limitation makes these modalities less suitable for complex video games requiring the management of multiple, rapid commands, such as action or shooter games.
}
Another possible solution is using alternative controllers, which offer different button layouts compared to standard game controllers~\cite{martinez2024playing,yuan2011game}.
For example, some accessible game controllers are designed to be operated with a single hand~\cite{xboxAdaptiveJoystick} or placed on a flat surface~\cite{liteSE, xboxAdaptiveController, ps5accessController}.
Others offer full personalization of the button layout through a modular design~\cite{xboxAdaptiveController,flexController,ps5accessController,proteusController}.
\revisionCHI{One limitation of alternative controllers is that they do not alter the underlying game mechanics, so players may still face challenges related to the high number of inputs, their coordination, or the quick reaction times inherent to the game design~\cite{ahmetovic2026shared}.}

These techniques can also be combined, but can still fall short.
In these cases, people with upper-limb impairments may resort to shared control solutions~\cite{motahar2022review, zhou2024coplayingvr, bhardwaj2025metatwin, ahmetovic2026shared,cimolino2021role}, delegating some game actions to another player~\cite{cimolino2021role,martinez2024playing,dalgleish2023can}.
The supported player is the \textit{pilot}, while the assisting player is the \textit{copilot}~\cite{xboxControllerAssist}.
When the pilot is supported by a \textit{human copilot}, shared control is referred to as \textit{human cooperation}; when there is a \textit{software copilot}, as \textit{partial automation}~\cite{ahmetovic2026shared}.

\subsection{Shared Control}
\label{ssec:shared-control}

After summarizing prior work on human cooperation (Section~\ref{sssec:related-hc}) and partial automation (Section~\ref{sssec:related-pa}), we analyze the main features of existing shared control systems (Section~\ref{sssec:related-sc-systems}), which form the basis for the design of \system{}.

\subsubsection{Human Cooperation}
\label{sssec:related-hc}

\revisionCHI{Human cooperation is adopted by players with disabilities to reduce the physical and temporal demands of gameplay.
Players typically request support when games require rapid inputs, precise movements, or multiple simultaneous commands, which they may struggle to execute consistently.
At the same time, they prefer to retain control over the core aspects of gameplay, delegating only secondary tasks to the copilot~\cite{ahmetovic2026shared}.}

Commercial implementations of human cooperation typically allow two players to control all game actions simultaneously, through two separate game controllers.
Both Microsoft's \textit{Xbox} and Sony's \textit{PlayStation} consoles provide integrated human cooperation solutions with the intent of improving game accessibility.
%Microsoft's \textit{Xbox Controller Assist}~\cite{xboxControllerAssist}, is available on both \textit{Xbox} consoles and \textit{Windows} PC, enabling the pairing of two \textit{Xbox} controllers for collaborative play.
%Similarly, \textit{Sony}'s \textit{Assist Controller}~\cite{ps5assistController} enables human cooperation on PS5 consoles.
In contrast, the \textit{Titan Two} adapter~\cite{titanTwo} is a third-party physical tool that
%, unlike the previous solutions,
also enables remote human cooperation.

Prior research has explored the application of human cooperation to video games~\cite{loparev2014introducing,sykownik2017exploring,rozendaal2010exploring,gonccalves2021exploring,ahmetovic2026shared}.
Most studies focus on analyzing the social benefits of collaborative play.
For instance, the \textit{WeGame} framework~\cite{loparev2014introducing} was used to demonstrate that human cooperation is generally a positive experience for the players involved, stimulating collaboration and encouraging thinking as a group.
Similarly, Sykownik et al.~\cite{sykownik2017exploring} found that, even when individual control is diminished, the resulting interdependence can induce enjoyment and social presence.
Rozendaal et al.~\cite{rozendaal2010exploring}, however, highlight that while shared control increases sociality, maintaining a sense of agency is essential for players, suggesting that action distribution should be carefully balanced to foster collaboration without undermining player autonomy.

Some studies also explored the potential of human cooperation as an accessibility tool~\cite{ahmetovic2026shared,gonccalves2021exploring}.
%The efficacy of human cooperation as an accessibility tool has been confirmed in previous works~\cite{ahmetovic2026shared,gonccalves2021exploring}.
Gonçalves et al.~\cite{gonccalves2021exploring} investigate how two players -- one with and one without visual impairments -- can play together through human cooperation, demonstrating that asymmetric roles during gameplay can result in an equitable gaming experience.
However, while human cooperation is an invaluable video game accessibility accommodation, the need for a human copilot represents a major limitation that can possibly be addressed through partial automation~\cite{ahmetovic2026shared}.
%For this reason, in this study, we explore shared control, which also includes partial automation, as an accessibility solution for video games.

\subsubsection{Partial Automation}
\label{sssec:related-pa}

Partial automation has been implemented in a few commercial video games, for example, for aim-assistance in first-person shooters~\cite{callOfDuty} or steering assistance in racing games~\cite{marioKart, forzaMotorsportAccessibility}.
In the \textit{Bayonetta} series~\cite{bayonetta, bayonetta2}, more advanced automation allows for delegating movement and targeting to the game, helping players focus solely on attacking.
In some of these cases, partial automation was explicitly designed as an accessibility support~\cite{forzaMotorsportAccessibility}.
However, in all cases, partial automation addresses specific game tasks and cannot be applied to all game actions.

Partial automation has also been explored in prior research~\cite{cechanowicz2014improving, cimolino2021role, cimolino2023automation, hwang2017game, medeiros2015developing}.
For example, \textit{Dino Dash} and \textit{Dozo Quest}~\cite{cimolino2021role} are exergames that allow players with spinal cord injuries to delegate a subset of controls to an AI agent based on their abilities.
\textit{Zac - O Esquilo}~\cite{medeiros2015developing}, instead, simplifies interaction to a single switch by automating movement direction, leaving the player with the only task of choosing when to move.
Game balancing techniques to dynamically adjust the provided level of assistance have also been investigated.
%Research has also investigated game balancing techniques to dynamically adjust the provided level of assistance to keep players engaged during gameplay.
Cechanowicz et al.~\cite{cechanowicz2014improving} studied game balancing in a multiplayer game, improving experience for novice players while offering a new challenge to more experienced ones.
Hwang et al.~\cite{hwang2017game} focused instead on tailoring the automation to the specific physical capabilities of the players in an exergame, aiming to maintain an engaging level of challenge while preserving the exercise's intended intensity.
%In order to maintain a high sense of perceived control and reduce frustration when interacting through a brain-computer interface, Hougaard et al.~\cite{hougaard2021willed} proposed instead an assistance mechanism that generates game actions on behalf of the player whenever they fail to execute an action or miss the required timing, effectively masking these errors.
%The impact of such assistance is further highlighted by Hougaard et al.~\cite{hougaard2021willed}, who demonstrated that system-generated game actions can actually increase perceived control and reduce frustration.

%Despite these benefits, the practical application of partial automation remains complex.
Previous papers also explore the limitations of partial automation.
Cimolino et al. investigated the problem of ``automation confusion''~\cite{cimolino2023automation}, which emerges when players struggle to understand which actions they control and which ones are automated.
Additionally, the partial automation solutions discussed above are all specifically designed and implemented for each considered video game, thus lacking generalizability to arbitrary third-party games, which limits their adoption as accessibility solutions.
Another limitation was observed by Ahmetovic et al.~\cite{ahmetovic2026shared}, who noted that substituting human cooperation with partial automation may reduce the social dimension of gaming.
% a strong interest in partial automation among players with disabilities, it lacks direct evidence from actual usage.
%As a result, the practical effectiveness of these systems as an alternative to human cooperation has yet to be evaluated through real-world studies. %gameplay experience.
%Another prior work also explores, through interviews and focus groups, the interest of gamers with disabilities in partial automation solutions~\cite{ahmetovic2026shared}.
%However, study participants had no actual experience with partial automation, and no empirical evaluation was conducted to determine the effectiveness of partial automation as an accessibility alternative for human cooperation.

Cimolino and Graham~\cite{cimolino2022two} analyze existing partial automation solutions and classify them according to a taxonomy spanning four dimensions: \textit{AI role} (\quots{who does what and when}), \textit{supervision} (\quots{whether actors correct each other}), \textit{influence} (\quots{how actors' actions influence each other}), and \textit{mediation} (\quots{how commands are unified}).
%Considering video games used for the experiments, Cimolino and Graham use ad hoc implementations of partial automation that either limit its application to specific aspects of the game or are designed primarily for research purposes.
%, like other works in this field, without investigating how partial automation can be achieved in commercial third-party games.
This taxonomy describes how existing partial automation approaches model the interplay between a human player and an assistive software agent.
However, it
%
%unlike our approach, Cimolino and Graham 
does not specifically address the impact of shared control, including both human cooperation and partial automation, on the gaming experience of players with disabilities.
%nor it explores design considerations for building shared control systems tailored to their needs.
%Additionally, they do not consider human cooperation systems nor compare them with partial automation approaches.
%
%Another difference concerns the games employed in the respective studies.
%Cimolino and Graham analyzed either partial automation features already available in commercial games -- typically limited to specific game actions and usually not customizable -- or ad-hoc games specifically developed to enable the use of partial automation.
%In contrast, our work introduces a research probe -- \system{} -- that enables both partial automation and human cooperation mechanisms to be applied to existing games, without requiring modifications to the original game design.
%
For this reason, in Section~\ref{sssec:related-sc-systems} we analyze existing shared control systems through a different lens, exploring design considerations for building a research probe that can be configured to function like other probes proposed in the literature.
%\system{} in such a way that it can support researchers to investigate how people with upper-limb impairments access video games in shared control.
%Based on this analysis, we design and implement a research probe -- \system{} -- that enables both human cooperation and partial automation in existing video games (Section~\ref{sec:system}) and we use it to investigate the identified research gaps (Section~\ref{sec:results-qualitative}). 
%to conduct a comparative evaluation of the experience of people with upper-limb impairments playing video games in human cooperation and partial automation (Section~\ref{sec:experiment}).
%From this analysis, we derive design recommendations for the development of future partial automation solutions (Section~\ref{ssec:discussion_design}).

\subsubsection{Features of Shared Control Systems}
\label{sssec:related-sc-systems}

Existing human cooperation and partial automation solutions share recurring design characteristics that determine how control is distributed between the pilot and the copilot.
In this section, we analyze those design characteristics (summarized in Table~\ref{tab:sc-features}) through the lens of providing all these functionalities in a single research probe to experiment with shared control in human cooperation and partial automation.
%: the number of players who can participate simultaneously, how \textit{game actions} are subdivided among them, how commands are merged when actions are shared, and whether the system supports third-party video games.
%Table~\ref{tab:sc-features} compares these features across the shared control systems surveyed in the previous two sections, also comparing them with \system{}, described in Section~\ref{sec:system}.

%Cimolino and Graham already defined a taxonomy to classify partial automation systems (Section~\ref{sssec:related-pa}).
%However, their taxonomy does not apply to human cooperation systems, as it considers dimensions such as \textit{AI role} and \textit{influence}, which are specific to partial automation.
%For this reason, here we focus instead on four main features of shared control systems that we identified as important for conducting experiments on shared control: the number of human players who can play together, how \textit{game actions} are subdivided between players, how players’ commands are merged when \textit{game actions} are shared, and whether the system supports third-party video games.

%We use this framework to conduct a comparative evaluation of the experience of people with upper-limb impairments playing video games in human cooperation and partial automation (Section~\ref{sec:experiment}). From this analysis, we derive design recommendations for future partial automation solutions (Section~\ref{ssec:discussion_design}).

\begin{table}[htb]
\centering
\caption{Main features of shared control systems in the state of the art. HC: Human Cooperation, PA: Partial Automation}
\label{tab:sc-features}
\begin{tabular}{lcccccc}
\toprule
 &
 HC &
 PA &
 \begin{tabular}[c]{@{}c@{}}Max human\\ players\end{tabular} &
 \begin{tabular}[c]{@{}c@{}}Shared/separated\\ game actions\end{tabular} &
 \begin{tabular}[c]{@{}c@{}}Merging policies\end{tabular} &
 \begin{tabular}[c]{@{}c@{}}Third-party\\ games support\end{tabular}
 \\
\midrule
Titan Two~\cite{titanTwo} & \checkgreen & \xred & 2 & Shared & Non-modifiable & \checkgreen \\
Gonçalves et al.~\cite{gonccalves2021exploring} & \checkgreen & \xred & 2 & Separated & Non-modifiable & \xred \\
Loparev et al.~\cite{loparev2014introducing} & \checkgreen & \xred & 4 & Both & 4 alternatives & \checkgreen \\
Xbox Controller Assist~\cite{xboxControllerAssist} & \checkgreen & \xred & 2 & Shared & Non-modifiable & \checkgreen \\
Rozendaal et al.~\cite{rozendaal2010exploring} & \checkgreen & \xred & 3 & Both & 3 alternatives & \xred \\
Assist Controller~\cite{ps5assistController} & \checkgreen & \xred & 2 & Shared & Non-modifiable & \checkgreen \\
Sykownik et at.~\cite{sykownik2017exploring} & \checkgreen & \xred & 2 & Both & 4 alternatives & \xred \\
%\midrule
Cechanowicz et al.~\cite{cechanowicz2014improving} & \xred & \checkgreen & 1 & Shared & Non-modifiable & \xred \\
Cimolino et al.~\cite{cimolino2021role} & \xred & \checkgreen & 1 & Separated & Non-modifiable & \xred \\
Cimolino et al.~\cite{cimolino2023automation} & \xred & \checkgreen & 1 & Both & Non-modifiable & \xred \\
%Hougaard et al.~\cite{hougaard2021willed} & \xred & \checkgreen & 1 & Shared & Non-modifiable & \xred \\
Hwang et al.~\cite{hwang2017game} & \xred & \checkgreen & 1 & Shared & Non-modifiable & \xred \\
Medeiros and Coutinho~\cite{medeiros2015developing} & \xred & \checkgreen & 1 & Shared & Non-modifiable & \xred \\
\midrule
\system{} (our solution) & \checkgreen & \checkgreen & $\infty$ & Both & Fully customizable & \checkgreen \\
\bottomrule
\end{tabular}
\end{table}

\paragraph{Number of Human Players}

The number of players refers to how many human participants can simultaneously take part in a shared control session.
At minimum, there is always one human player, the pilot.
In human cooperation systems, at least one human copilot is present by definition, while partial automation systems may operate with the pilot alone, assisted by a software agent.
Among commercial human cooperation systems~\cite{xboxControllerAssist, ps5assistController, titanTwo}, the maximum supported number of players is two.
Prior research on human cooperation has explored broader configurations, with some studies involving up to three or four simultaneous players~\cite{rozendaal2010exploring, loparev2014introducing}.
In contrast, both commercial and research implementations of partial automation have consistently been limited to a single human player.
None of the analyzed systems allows for the cooperation of the pilot with both a 
human and software copilot at the same time, preventing the study of their combined 
effects on gameplay experience.

\paragraph{Game Actions Subdivision}

Most of the analyzed systems allow all players to control all game actions.
This configuration is adopted, for example, by commercial solutions such as Microsoft's \textit{Xbox Controller Assist} and the \textit{Titan Two}~\cite{titanTwo}, where the inputs of multiple controllers are merged and sent to the game.
Some studies explored configurations in which each player controls an exclusive subset of game actions.
For example, Gonçalves et al.~\cite{gonccalves2021exploring} assigned different actions to each player so that they had to cooperate to play the game.
Other works investigated multiple configurations.
For example, Sykownik et al.~\cite{sykownik2017exploring} compared two different scenarios where players could either issue commands independently or jointly control the same actions.
Rozendaal et al.~\cite{rozendaal2010exploring} also studied multiple configurations, 
including one in which only a single action was shared among all players, 
while the remaining actions were controlled exclusively.
%
%We note that in most prior works multiple players can have control of a same game action.
%Such configurations, as we show in the following (Section~\ref{ssec:confusion}), can lead to coordination issues during play (\textit{e.g.}, players not knowing who should intervene in a given moment).
%Therefore, when playing in human cooperation, users typically agree on how to divide control responsibilities before starting to play~\cite{ahmetovic2026shared}.
%Prior work in partial automation has likewise shown that overlapping control of the same game actions should be avoided, as it may lead to automation confusion~\cite{cimolino2023automation, parasuraman1997humans}.

\paragraph{Merging Policies}

When a single game action is under the control of more than one player, there is a need to define a procedure to combine their inputs.
We refer to this procedure as \textbf{merging policy}.
All commercial human cooperation systems~\cite{xboxControllerAssist,ps5assistController} adopt the same merging policy: for binary inputs (\textit{e.g.}, buttons), if at least one player performs the action, the action is executed; for continuous inputs (\textit{e.g.}, joysticks), the controls from all players are summed.
For example, 
%using \textit{Xbox Controller Assist}~\cite{xboxControllerAssist}
in a platform game, if both players move slightly to the left, the character will quickly move in that direction.
Other merging policies are possible as well.
%For example, in Cechanowicz et al.'s system~\cite{cechanowicz2014improving}, the partial automation mechanism adds a correction factor to the player's steering input, so the final input sent to the game is a combination of the human's command and the AI's adjustment.
For example, one configuration in Sykownik et al.’s \textit{Shairit} game~\cite{sykownik2017exploring} averages the inputs for the movement game action.
So, the character moves in a certain direction at maximum speed only when all players provide input in that same direction.
%In Hwang et al.'s \textit{Gekku Aim} game~\cite{hwang2017game}, instead, the aiming direction is selected between the input specified by the human player and the optimal direction computed by a partial automation mechanism.
%
Finally, some systems support multiple merging policies, making it possible to select one from a fixed set of alternatives~\cite{loparev2014introducing, rozendaal2010exploring}.
%By testing multiple alternatives, 
These works demonstrate that varying how inputs are merged alters the players' social interaction, perceived autonomy, and overall engagement during collaborative play.
%For example, Loparev et al.~\cite{loparev2014introducing}, Rozendaal et al.~\cite{rozendaal2010exploring}, and Sykownik et al. tested multiple alternative policies.

\paragraph{Support for Third-Party Video Games}

%In order to study partial automation and human cooperation in real-world scenarios, \system{} needs to support third-party video games.

In the context of video games, few shared control systems are compatible with existing third-party titles.
There are some commercial systems~\cite{xboxControllerAssist, ps5assistController, titanTwo,loparev2014introducing} that support them and are used by people with disabilities~\cite{ahmetovic2026shared}.
However, these systems only support human cooperation.
Regarding partial automation, prior work has studied how it could be applied for automating specific video games, which were often developed for research purposes only~\cite{cimolino2021role, cimolino2023automation, cechanowicz2014improving, hwang2017game, medeiros2015developing}.
Some commercial video games also implement partial automation, but limited to some specific game tasks in the considered game and therefore not generalizable (Section~\ref{sssec:related-pa}).

\section{The \systemFullNameTitle{}}
\label{sec:system}

%Sergio: rifrasato per accorciare
%The analysis of existing shared control systems in Section~\ref{sssec:related-sc-systems} shows that no single system supports all the features exposed by shared control systems.
%As summarized in Table~\ref{tab:sc-features}, no solution supports both human cooperation and partial automation, and most also lack support for third-party games and offer fixed merging policies.
%To address these gaps, we developed the \systemFullName{} as a research probe that can reproduce all the interaction modalities offered by existing shared control systems.

%In this section, we describe how \system{} implements these features.
%We also describe how it supports input remapping, a widely adopted accessibility feature (Section~\ref{ssec:related-accessibility}) that we considered essential for enabling players with upper-limb impairments to participate in our study.
%Details on how \system{} was implemented as a standalone application are available in Appendix~\ref{appendix:implementation}.
%The source code is publicly available\footref{fn:source-code}.

%As summarized in Table~\ref{tab:sc-features}, no solution supports both human cooperation and partial automation.
%To ease the experimental evaluation in this field, we developed \system{}.
This section describes the framework features implemented in \system{} and used in our study.
Implementation details are reported in Appendix~\ref{appendix:implementation}, and the source code is publicly available\footref{fn:source-code}.

%TC:ignore
\begin{comment}
\revisionCHI{
%Contrary to prior research, which explores partial automation implementations in ad hoc video games~\cite{cimolino2021role,hwang2017game,cimolino2023automation,medeiros2015developing},
We developed the \systemFullName{} as a research probe for investigating shared control in real-world scenarios using existing commercial third-party video games.
Specifically, since we aim to explore how user experience differs between human cooperation and partial automation, \system{} was built to support all shared control interaction modalities already presented in prior literature~\cite{loparev2014introducing, sykownik2017exploring, rozendaal2010exploring, gonccalves2021exploring, ahmetovic2026shared, medeiros2015developing, cimolino2021role, cechanowicz2014improving}.
In this section, we identify the key requirements of \system{} by analyzing prior implementations of shared control, and we describe the main features of the framework.
Throughout the analysis, we also explain how \system{} can support all the partial automation scenarios captured by Cimolino and Graham's taxonomy~\cite{cimolino2022two}, showing that our framework enables rapid experimentation of different partial automation configurations.
Details on how \system{} was implemented as a standalone application are available in Appendix~\ref{appendix:implementation}, while the source code is publicly available\footnote{\url{https://anonymous.4open.science/r/GamePals-B88B}}.}
\end{comment}
%TC:endignore

\subsection{\revisionCHI{Seamless Integration of Human Cooperation and Partial Automation}}

% Human cooperation
% \revisionCHI{Commercial human cooperation systems~\cite{xboxControllerAssist,ps5assistController,titanTwo} support third-party video games, but limit the number of players to a human pilot and a human copilot.
% Instead, prior work in the literature explored more flexible human cooperation modalities, also applied to more than two players~\cite{rozendaal2010exploring,loparev2014introducing}.}

% % Partial automation
% \revisionCHI{Conversely, for partial automation, there are no general commercial solutions applicable to third-party video games.
% Instead, some video games directly implement partial automation for specific tasks~\cite{callOfDuty,forzaMotorsportAccessibility,bayonetta,bayonetta2,marioKart,kingdomComeDeliverance}.
% For example, in Mario Kart~\cite{marioKart}, the acceleration can be entirely delegated to the software copilot, while the human copilot controls all other game actions.
% Partial automation is also investigated in prior research~\cite{cimolino2021role,gonccalves2021exploring}.
% However, these solutions are not applicable to third-party games, as they are typically used with specifically crafted games.}

% Gamepals
Unlike existing shared control systems, \system{} supports both human cooperation and partial automation, with all possible player configurations: one or more human players and zero or more software players.
%Each human player can act either as a \textbf{human pilot} or a \textbf{human copilot}, while the software player acts as a \textbf{software copilot} that supports the other players by controlling some \textit{game actions}.
Human cooperation is supported out of the box for any video game, while partial automation requires game-specific adaptations (Section~\ref{ssec:third-party}).

\begin{figure}[htb]
    \centering
    \begin{subfigure}{0.49\linewidth}
        \centering
        \includegraphics[height=.4\textheight]{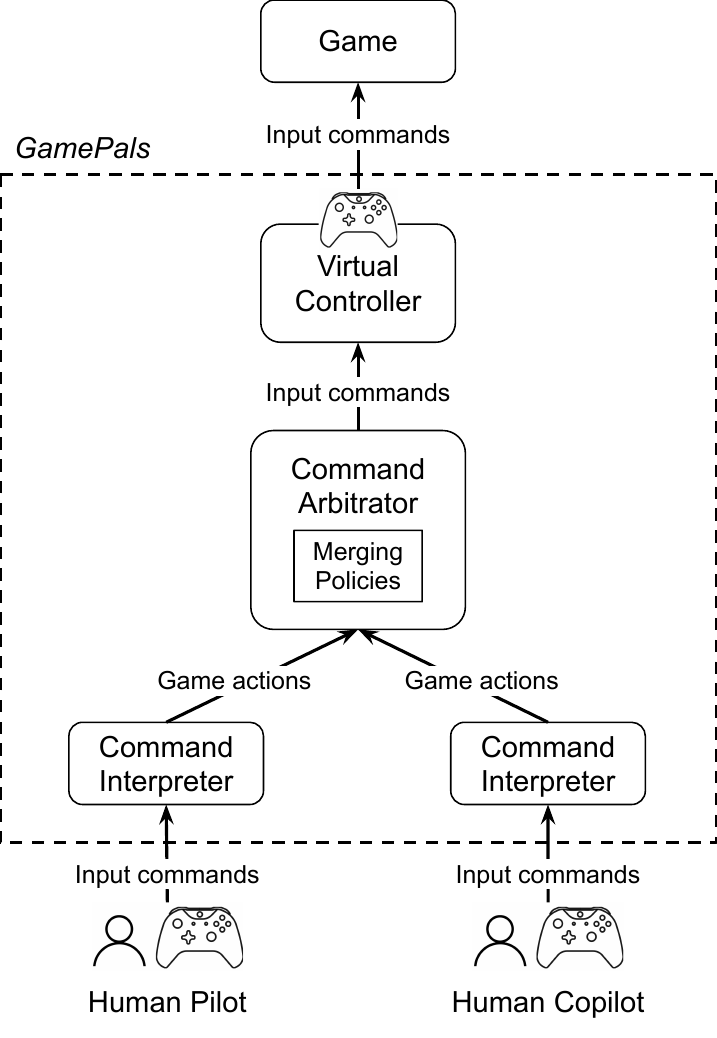}
        \caption{\textcolor{red}{Human cooperation}}
        \label{fig:sc-architecture-hc}
        \Description{A diagram showing the data flow of GamePals in a human cooperation setup. Two players are present, depicted using a standard game controller: a human pilot and a human copilot.
        Both players produce, using their controllers, input commands. Input commands are sent to a command interpreter, which maps them into game actions.
        The game actions are sent to a command arbitrator, which merges the two streams by applying merging policies. After merging, it maps the game actions back into input commands, which are sent to the game.
        A large box encloses the part of the diagram that constitutes the GamePals framework. The box includes the two command interpreters, the command arbitrator and the virtual controller.}
    \end{subfigure}
    \begin{subfigure}{0.49\linewidth}
        \centering
        \includegraphics[height=.4\textheight]{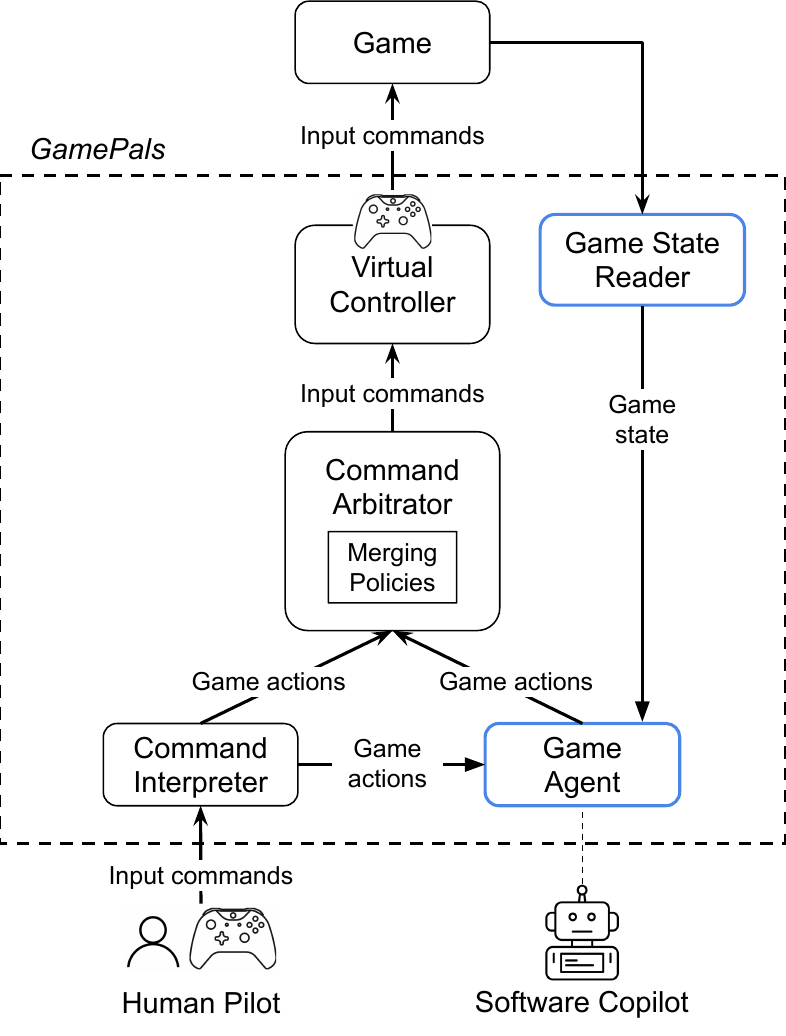}
        \caption{\textcolor{blue}{Partial automation}}
        \label{fig:sc-architecture-pa}
        \Description{A diagram showing the data flow of GamePals in a partial automation setup. Two players are present: a human pilot and a software copilot. The human pilot is depicted with a standard Xbox controller; the software copilot is depicted using a robot icon and a rectangle representing the game agent.
        The human pilot produces, using their controller, input commands. Input commands are sent to a command interpreter, which maps them into game actions.
        The game actions are sent both to the command arbitrator and the game agent.
        The game agent receives the game state from a game state reader, which is represented as a module close to the game. It then produces game actions, which are also sent to the command arbitrator.
        The command arbitrator thus receives game actions from both the command interpreter and the game agent. It merges the two streams by applying merging policies. After merging, it maps the game actions back into input commands, which are sent to the game.
        A large box encloses the part of the diagram that constitutes the GamePals framework. The box includes the two command interpreters, the command arbitrator, the virtual controller, the game state reader and the game agent.}
    \end{subfigure}
    \caption{\revisionCHI{\systemFullName{}'s data flow}}
    \label{fig:sc-architecture}
\end{figure}

In our study (Section~\ref{sec:experiment}), we use configurations involving one pilot and one copilot, as this is the only setup supported in commercial shared control systems.
%represents the most common setup in human cooperation, where a single copilot supports the pilot during gameplay~\cite{ahmetovic2026shared}.
One configuration implements human cooperation, where the copilot is another human player (Fig.~\ref{fig:sc-architecture-hc}), while the other implements partial automation, where the copilot is a software agent (Fig.~\ref{fig:sc-architecture-pa}).

\subsection{\revisionCHI{User-specific Input Remapping}}
\label{ssec:input-remapping}

\revisionCHI{
%A game controller generates \textbf{input commands} when a user interacts with its elements (\textit{i.e.}, buttons, triggers, analog sticks) and sends them to a game.
%When \textit{input commands} are received, video games trigger the corresponding \textbf{game actions}.
%Examples of \textit{game actions} include shooting and jumping in first-person shooters or steering in driving games.
%Each game defines a default mapping between each \textit{game action} and an \textit{input command}.
%For instance, in a racing game played with an \textit{Xbox} controller~\cite{xboxController}, the downshift and upshift \textit{game actions} may be mapped to the left and right bumpers, the acceleration to the right trigger, and the steering to the left analog stick.
Many games allow the user to personalize how the controller \textbf{input commands} are mapped to the \textbf{game actions}.
This remapping is often used as an accessibility accommodation~\cite{ahmetovic2026shared, brown2021designing, ahmetovic2021replay, playAbility, voiceAttack, joyToKey, martinez2024playing}.
For example, in a racing game%
%played with an \textit{Xbox} controller~\cite{xboxController}
, acceleration may be mapped by default to the right trigger, and steering to the left analog stick.
In such a case, a user who has difficulties in controlling their left hand could remap steering to the right analog stick.
%For example, in the previous case, the steering could be remapped to the right analog stick by a user who has difficulties in controlling their left hand.
}
\revisionCHI{Conversely, existing commercial human cooperation systems~\cite{xboxControllerAssist, ps5assistController, titanTwo} require all players to control \textit{game actions} using the same \textit{input commands}.
This means that pilot and copilot cannot remap \textit{input commands} to \textit{game actions} differently, even if they have different preferences and accessibility needs.
To address this limitation, \system{} was designed to allow each human player to personalize the mapping between \textit{game actions} assigned to them and \textit{input commands}.}

\revisionCHI{Input remapping is implemented by the \textbf{command interpreter} component (Fig.~\ref{fig:sc-architecture}), which intercepts each human player's \textit{input commands}, maps them to the \textit{game actions} specified for that player, and forwards the result to the \textbf{command arbitrator} (Section~\ref{ssec:merging}).}

\subsection{\revisionCHI{\textit{Game Actions} Subdivision}}
\label{ssec:game-action-subdivision}

% \revisionCHI{In Cimolino and Graham's taxonomy~\cite{cimolino2022two}, the \quots{AI role} dimension (Section~\ref{ssec:related-pa}) defines which player controls which \textit{game action}.
% Commercial human cooperation systems~\cite{xboxControllerAssist,ps5assistController,titanTwo} typically enable all players to control all \textit{game actions}.
% This configuration maps onto the Cimolino and Graham's \quots{reciprocal} scenario.
% Instead, if only some of the \textit{game actions} are shared, four scenarios are possible (\quots{supportive}, \quots{delegated}, \quots{cooptable}, or \quots{complementary}), which differ based on who is the primary controller of the overlapping commands.
% %
% However, overlapping commands can lead to confusing situations during play (\textit{e.g.}, players not knowing who should intervene in a given moment).
% Therefore, users typically agree on how to divide control responsibilities before starting to play~\cite{ahmetovic2026shared}.
% Prior work in partial automation has likewise shown that overlapping control of the same \textit{game actions} should be avoided, as it may lead to automation confusion~\cite{cimolino2023automation,parasuraman1997humans}.}

\system{} allows specifying which \textit{game actions} are controlled by each player.
One possibility is for each player to control all \textit{game actions}.
Alternatively, each \textit{game action} can be assigned to a different player, avoiding situations in which two players control the same \textit{game action}.
These are the most common setups in the existing systems (Section~\ref{sssec:related-sc-systems}).
Finally, hybrid configurations allow each player to control some \textit{game actions} exclusively, while sharing others.

% \system{} supports all scenarios defined in Cimolino and Graham's taxonomy by allowing specifying which \textit{game actions} are controlled by each player.
% One possibility is that each player controls all \textit{game actions}.
% Alternatively, each \textit{game action} can be assigned to a different player, thereby avoiding situations in which two players control the same \textit{game action}.
% This second scenario, which is not considered in the taxonomy proposed by Cimolino and Graham, is indeed relevant because it represents a typical human cooperation setup~\cite{ahmetovic2026shared,gonccalves2021exploring}.
% Finally, hybrid configurations are also possible, in which each player controls some \textit{game actions} exclusively, while others are shared.

For human players, \textit{game actions} subdivision is implemented by the \textit{command interpreter} component (Fig.~\ref{fig:sc-architecture}), along with input remapping (Section~\ref{ssec:input-remapping}).
When an \textit{input command} is provided, the command interpreter forwards the associated \textit{game action} to the \textit{command arbitrator} component.
In case an \textit{input command} is not associated with any \textit{game action} for a specific player, it is filtered out.
For software players, instead, each \textit{game agent} (Section~\ref{ssec:third-party}) generates only the \textit{game actions} assigned to it and directly forwards them to the \textit{command arbitrator}.

\subsection{\textit{Game Actions} Merging}
\label{ssec:merging}

\system{} has a modular architecture that makes it possible to implement arbitrary \textbf{merging policies}, choosing a different one for each \textit{game action}.
%
% Architecture
\textit{Game actions} merging is implemented by the \textit{command arbitrator} module (Fig.~\ref{fig:sc-architecture}), that takes the \textit{game actions} received by each \textit{command interpreter} and \textit{game agent} (Section~\ref{ssec:third-party}) and merges them by applying the \textit{merging policy} configured for that \textit{game action}.
The \textit{merging policy} can also take into account the player’s role (\textit{i.e.}, pilot or copilot), allowing different behaviors depending on who issues the command.
For example, a \textit{merging policy} may select the pilot’s command in the event of conflicting \textit{game actions}.
The resulting \textit{game action} is converted into the \textit{input command} that the game associates with that \textit{game action}, and it is forwarded to the \textit{virtual controller} (Section~\ref{ssec:third-party}).

% Study
To conduct the experiments presented in this paper, we adopted the same \textit{merging policies} implemented by commercial human cooperation implementations~\cite{xboxControllerAssist, ps5assistController, titanTwo}.
Additional details about them are provided in Appendix~\ref{appendix:policies}.
%\textit{Selected mediation} scenarios can also be implemented with more complex \textit{merging policies} that account for players' roles.
%For example, a \textit{merging policy} may select the pilot’s command in the event of conflicting \textit{game actions}.
%
% Cimolino
% \textit{Merging policies} can also implement all the scenarios belonging to the Cimolino and Graham's \quots{supervision} dimension, which \quots{specifies how actors monitor and correct each other}.
% In our study, we used \quots{unsupervised supervision}, meaning that \quots{neither actor supervises the other}.
% Other scenarios, like \quots{by AI supervision}, can be defined through a \textit{merging policy} that selectively overrides human input under specific conditions.

\subsection{\revisionCHI{Integration With Third-Party Video Games}}
\label{ssec:third-party}

\system{} integrates with existing third-party games by sending the merged \textit{game actions} to a virtual controller, which the game interprets as a standard physical controller.
This enables human cooperation in any controller-based game\footnote{Certain video games only work with mouse and keyboard and are currently not supported.}.
Partial automation additionally requires a \textbf{game adaptation}, which is specific to each game and composed of a \textbf{game state reader} and a \textbf{game agent}: the former extracts the current \textit{game state}, while the latter receives the \textit{game state}, generates the copilot’s actions, and forwards them to the command arbitrator (Fig.~\ref{fig:sc-architecture-pa}).
In our study, we implemented this adaptation for \textit{Rocket League} (Section~\ref{ssec:apparatus}) through modding\footnote{Modding is \quots{the act of changing a game, usually through computer programming, with software tools that are not part of the game}~\cite[p. 1250]{poor2014computer}.}.

% \textit{Game agents} are particularly important for \system{} to support the \quots{influence} dimension, as defined by Cimolino and Graham~\cite{cimolino2022two}.
% This dimension \quots{captures how an actor chooses its own actions in response to those of the other actor}.
% In this study, we used the \quots{independent} modality when playing in partial automation, where \quots{neither actor influences the other}.
% Other modalities can also be configured.
% Indeed, by providing the \textit{game agent} with the \textit{game actions} performed by the human player, \system{} can support \quots{interpreted influence} (\quots{the AI’s actions are influenced by the human’s actions}).
% \system{} can also be easily extended to support the \quots{guided} modality (\quots{the human's actions are influenced by the AI's actions}) by displaying the \textit{game actions} entered by the \textit{game agent} to the human player.
% Thus, the \quots{codependent} modality, which combines \quots{interpreted} and \quots{guided} modalities, is also possible. 
% However, while \system{} supports all these interaction modalities, previous literature has observed that player interaction in shared control can be complex, involving vocal communication, advanced visual feedback, and intent understanding~\cite{ahmetovic2026shared}.
% These advanced forms of interaction are not yet implemented in \system{} and will be the subject of future research.

\section{Experimental Methodology}
\label{sec:experiment}

Through a study with $13$ representative participants, we investigate how people with upper-limb impairments play in human cooperation and in partial automation, collecting their subjective feedback on the experience.
%with these two assistive technologies.
The study was approved by the Ethics Committee of \anon[our institution]{the University of Milan (opinion no. 48/25)}\footnote{\revisionCHI{Planning, conduct, and reporting of the work are consistent with the general principles enunciated in~\cite{acmetics}.}}, it lasted approximately one hour for each participant and was supervised by at least one member of the research team.
%throughout its entire duration.
%The complete study protocol is available in Appendix~\ref{appendix:protocol}.

\subsection{Game Stimulus and Conditions}
\label{ssec:stimulus}

\begin{figure}[htb]
    \centering
    \includegraphics[width=0.5\linewidth]{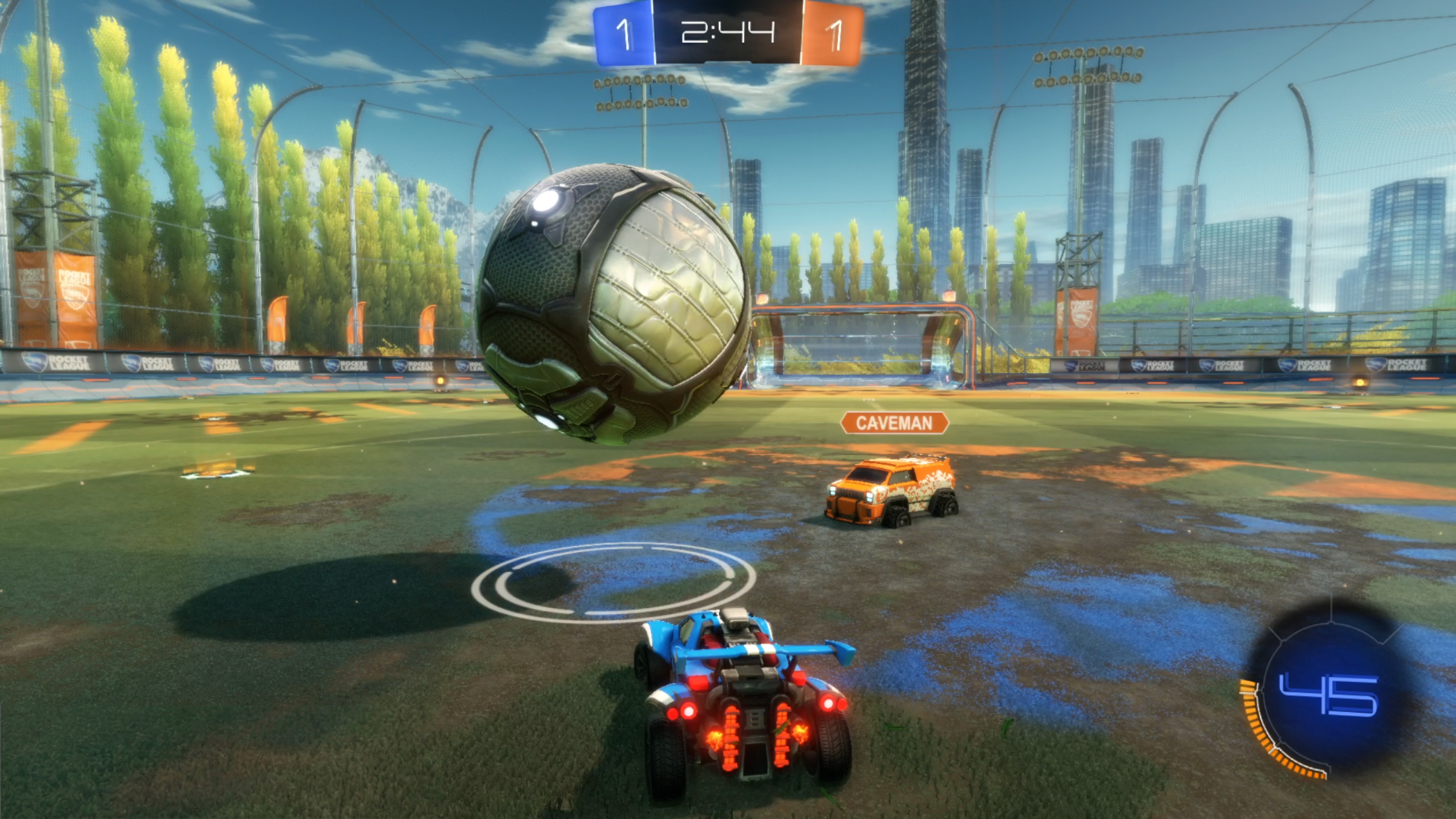}
    \caption{In-game screenshot from \textit{Rocket League}}
    \label{fig:rocketLeague}
    \Description{Screenshot of a Rocket League match. The player controls a blue car. The opponent controls an orange car. Both players are driving towards the ball, which is in the air in the center of the field.}
\end{figure}

The video game used as stimulus for the study is \textit{Rocket League}~\cite{rocketLeague} (Fig.~\ref{fig:rocketLeague}), a team ball game (similar to soccer) in which each player controls a vehicle and aims to score with a large ball (roughly 3 times taller than the vehicles) into the opponent team's goal.
Each match lasts $5$ minutes, and the team that scores more goals by the end wins.
Players control their vehicle through a variety of actions, including accelerating, braking, steering, jumping, handbraking, and boosting.
The handbrake enables tight turns, while the boost provides \revisionCHI{short bursts of acceleration limited by the available boost fuel, which can be recharged by moving over the charging pads.
Jumps can be used to reach the ball high in the air, to move faster, and to quickly rotate in mid-air.}
\revisionCHI{\textit{Rocket League} was selected because it has a clearly defined objective, allowing participants to quickly understand the game within the limited time frame of the experiment, without extensive training.}
Additionally, the game's Pan European Game Information (PEGI) rating of 3~\cite{pegi} ensures its suitability for a diverse audience.

%The video game used as stimulus for the study is \textit{Rocket League}~\cite{rocketLeague} (Fig.~\ref{fig:rocketLeague}), a real-time vehicle soccer game involving steering, acceleration, braking, jumping, boosting, and handbraking.
%We selected it because it has a clear objective and supports rapid multi-input gameplay.

During the study, participants were asked to play two \textit{Rocket League} matches: one in human cooperation, and one in partial automation.
To avoid influences from other players, we chose to conduct single-player 1v1 matches, in which the participant plays against a bot (a software agent controlled by the game).

\subsection{Apparatus}
\label{ssec:apparatus}

The experimental apparatus consists of the \systemFullName{} (Section~\ref{sec:system}), the game stimulus -- \textit{Rocket League} -- (Section~\ref{ssec:stimulus}), and the corresponding \textit{game adaptation} composed of a \textit{game state reader} and a \textit{game agent}.

For the \textit{game state reader}, we used the \textit{BakkesMod} modding framework~\cite{bakkesMod} to develop a plugin for \textit{Rocket League} that reads the entire \textit{game state} and transmits it to \system{}.
The \textit{game agent} was developed by adapting a pre-trained bot available online\footnote{We adapted the \textit{Nexto bot}, which is a freely available high-performing \textit{Rocket League} bot: \url{https://github.com/Rolv-Arild/Necto}} so that it could control only the \textit{game actions} assigned to it.

The experiment ran on a laptop\footnote{Lenovo ThinkBook 16p Gen5 with Windows 11, Intel Core i7-14650HX CPU, 32 GB of RAM, and NVIDIA GeForce RTX 4060 Mobile graphics card.}.
Participants could choose to use either a standard \textit{Xbox} controller or an \textit{Xbox Adaptive Controller}~\cite{xboxAdaptiveController}, optionally combined with the \textit{Logitech Adaptive Gaming Kit}~\cite{logitechAdaptiveGamingKit} and the \textit{Logitech Extreme 3D Pro Joystick}~\cite{logitechJoystick}.
All \textit{game actions} could be remapped to any available button on the chosen controller (Section~\ref{ssec:input-remapping}).
More details on the game and on its adaptation are provided in Appendix~\ref{appendix:game-agent}.

\subsection{Protocol}
\label{ssec:protocol}
After a brief introduction to the study, its goals, and its structure, the supervisor collected the participant's consent and demographic data.
Then, the supervisor explained the game mechanics, supported by a pre-recorded gameplay video.

During the following configuration phase, the participant 
%was supported by the supervisor in selecting the game configuration.
%This included 
decided, for each \textit{game action}, whether to control it directly, assign it to the copilot, or share it.
%if the participant would control it directly, assign it to the copilot, or share it with the copilot.
\revisionCHI{Participants freely chose according to their preferences. The supervisor provided guidance by explaining the different \textit{game actions}, offering suggestions, and clarifying the potential implications of each choice.}
In addition, the participant selected their preferred controller and assigned each \textit{game action} they decided to control to an \textit{input command} on that controller.
The chosen configuration was then tested in-game, with the supervisor acting as copilot.
This configuration and testing procedure were repeated until the participant found the configuration suitable.
Once defined, the configuration remained the same throughout the study.

After configuring the system, participants were asked to play two sessions, each approximately $5$ minutes long: one in human cooperation and the other in partial automation.
To avoid order bias, the sequence of the two sessions was counterbalanced using a Latin Square design.
During the sessions, screen output, ambient audio, and \textit{input command} logs were recorded with the participant's consent, to support the analysis.
\revisionCHI{
%Participants' \textit{input command} logs were also saved.
Participants were encouraged to think-aloud~\cite{lazar2017research}, expressing their reasoning in real time and providing insights into their experience.}

Finally, participants took part in a semi-structured interview~\cite{lazar2017research} reflecting on their experiences with both shared-control modalities. 
The interview questions were informed by prior work on shared-control assistance~\cite{cimolino2022two, cimolino2023automation,gonccalves2021exploring} and covered participants' modality preferences, perceived control, moments of frustration, interpretability of the software copilot's actions, and desired improvements. 
Follow-up questions were asked to clarify or expand on participants' responses. 
The full protocol, including the complete interview guide, is reported in Appendix~\ref{appendix:protocol}.

\subsection{Participants}
\label{ssec:participants}

Participants were recruited through local associations of people with motor impairments: \anon{Spazio Vita\footnote{\url{https://spaziovitaniguarda.it}} of Niguarda Hospital in Milan - Italy, and Milan and Monza's branches of UILDM\footnote{Unione Italiana Lotta Distrofia Muscolare: \url{https://www.uildm.org}}}.
%These associations also provided locations for the experiments.
%
The following inclusion criteria were adopted:
\begin{itemize}
    \item Being at least 13 years old.
    \item Being able to speak \anon[languages spoken by the authors]{Italian}.
    \item Having an upper-limb impairment that limits ability to use standard video game controllers.
\end{itemize}
The only exclusion criterion was the presence of multiple disabilities, as %they could be a 
potential confounding factor for the analysis. 
%Prior experience with \textit{Rocket League} was not considered a confounding factor, as we expected it to influence both sessions in a similar way.

Following these criteria, $13$ participants were recruited.
One additional participant, who did not complete either session due to motion sickness, was excluded from the analysis. 
%They were therefore excluded because they did not complete either session. 
Participants' demographic data is reported in Table~\ref{tab:demographic-data}.
%did not complete either session.
%For this reason, they were excluded from the analysis.
%The demographic profiles of the participants are reported in Table~\ref{tab:demographic-data}.

\begin{table}[htb]
\caption{Participants' Demographic Data. RL = Rocket League, HC = Human Cooperation}
\label{tab:demographic-data}
\centering
\footnotesize
\begin{tabular}{|l|l|l|l|l|l|l|l|l|l|}
\hline
\multirow{2}{*}{\textbf{ID}} &
\multirow{2}{*}{\textbf{Age}} &
\multirow{2}{*}{\textbf{Gender}} &
\multicolumn{2}{c|}{\textbf{Disability}} &   
\multicolumn{5}{c|}{\textbf{Gameplay}}
\\ \cline{4-5}\cline{6-10}
 &  &  &  \textbf{Type} &   
\textbf{Onset} &   
\textbf{Freq.} &   
\textbf{Difficulty} & \textbf{In RL} & \textbf{In HC} & \textbf{Platform} \\ \hline \hline
\ParticipantTwo{} &
  19-28 &
  M &
  SMA (Type 2) &
  18 months &
  Weekly &
  Quite a bit (4) &
  No & No &
  PC \\
\hline
\ParticipantThree{} &
  19-28 &
  M &
  Duchenne muscular dystrophy &
  12 years &
  Daily &
  A little (2) &
  Yes & No &
  PS \\
\hline
\ParticipantFour{} &
  19-28 &
  F &
  %\begin{tabular}[c]{@{}c@{}}Muscular dystrophy\\(Merosin-Deficient)\end{tabular} &
  Merosin deficiency &
  Birth &
  Weekly &
  A lot (5) &
  No & No &
  %Smartphone \\
  Mobile \\
\hline
\ParticipantFive{} &
  39-50 &
  M &
  %\begin{tabular}[c]{@{}c@{}}Muscular dystrophy\\(Becker)\end{tabular} &
  Becker muscular dystrophy &
  12 years &
  Daily &
  Moderately (3) &
  Yes & No &
  PS5 \\
\hline
\ParticipantSix{} &
  19-28 &
  M &
  Duchenne muscular dystrophy &
  Birth &
  Weekly &
  A little (2) &
  Yes & No &
  PC \\
\hline
\ParticipantSeven{} &
  51-65 &
  F &
  Amputee (hands and legs) &
  43 years &
  Daily &
  Quite a bit (4) &
  No & No &
  %\begin{tabular}[c]{@{}c@{}}Smartphone,\\ Tablet\end{tabular} \\
  Mobile \\
\hline
\ParticipantEight{} &
  19-28 &
  M &
  Duchenne muscular dystrophy &
  4/5 years &
  Weekly &
  Moderately (3)  &
  Yes & No &
  %PS, Smartphone \\
  PC, Mobile \\
\hline
\ParticipantNine{} &
  29-38 &
  F &
  Tetraplegia &
  Birth &
  Daily &
  Quite a bit (4) &
  No & No &
  %\begin{tabular}[c]{@{}c@{}}Smartphone,\\ Tablet\end{tabular} \\
  Mobile \\
\hline
\ParticipantTen{} &
  66-80 &
  M &
  %\begin{tabular}[c]{@{}c@{}}Muscular dystrophy\\(Amyotrophy)\end{tabular} &
  Amyotrophy &
  3 months &
  Weekly &
  Quite a bit (4) &
  No & No &
  %Smartphone \\
  Mobile \\
\hline
\ParticipantEleven{} &
  19-28 &
  M &
  %\begin{tabular}[c]{@{}c@{}}Muscular dystrophy\\(FSHD)\end{tabular} &
  FSHD &
  4/5 years &
  Daily &
  Not at all (1) &
  No & No &
  PC \\
\hline
\ParticipantTwelve{} &
  19-28 &
  M &
  Spastic Tetraparesis &
  Birth &
  Daily &
  Moderately (3) &
  Yes & Yes &
  Switch, PS5 \\
\hline
\ParticipantFifteen{} &
  39-50 &
  F &
  Left pediatric tetraparesis &
  Birth &
  Daily &
  Quite a bit (4) &
  No & Yes &
  %Smartphone \\
  Mobile \\
\hline
\ParticipantSixteen{} &
  19-28 &
  F &
  Reduced arm mobility &
  Birth &
  Daily &
  A little (2) &
  No & Yes &
  %Tablet \\ \hline
  Mobile \\ \hline \hline
\textbf{Copilot} &
  19-28 &
  M &
  None &
  None &
  Weekly &
  Not at all (1) &
  Yes & No &
  PC, Switch \\ \hline
\end{tabular}%
\Description{A table showing the demographic data of the participants. Each row corresponds to one of the participants. Columns correspond to their general demographic information, details about their disability, and information about their experience with gaming, with the Rocket League game, and with human cooperation.
The last row of the table corresponds to the human copilot involved in the experiments, for which the same information as the other participants is reported.
}
\end{table}

%Eight participants identified as male and five as female.
%The most represented age group was 19 to 28 (8 participants), followed by the age group 39 to 50 (2 participants).
%Other age groups had 1 participant each.
%All participants play video games at least once a week, most daily (8 participants).
%Only
\ParticipantEleven{} reported having no difficulty playing the video games they usually play.
All others reported having at least \quots{a little} difficulty (2 out of 5), with half of the participants reporting at least \quots{quite a bit} (4 out of 5).
%The most used gaming platform is the smartphone (6 participants), followed by \textit{PlayStation} (4 participants), PC and tablet (3 participants each), and \textit{Nintendo Switch} (1 participant).
%For seven participants, the choice was motivated by the platform’s ease of access (\ParticipantFour{}, \ParticipantSeven{}, \ParticipantNine{}, \ParticipantTen{}, \ParticipantTwelve{}, \ParticipantFifteen{}, \ParticipantSixteen{}).
%Other motivations included familiarity with it (\ParticipantFive{}, \ParticipantTwelve{}), the video games available for the platform (\ParticipantSix{}, \ParticipantTwelve{}), and the portability of the device (\ParticipantEight{}, \ParticipantTen{}).
Five participants use hardware or software tools to facilitate access to video games.
These include the \textit{Xbox Adaptive Controller}~\cite{xboxAdaptiveJoystick} (\ParticipantFifteen{}, \ParticipantSixteen{}), custom buttons and joysticks (\ParticipantTen{}), software for remapping controller buttons (\ParticipantTwo{}), and in-game accessibility options (\ParticipantTwelve{}).
Only \ParticipantTwelve{}, \ParticipantFifteen{}, and \ParticipantSixteen{} had previous experience with shared control.
Specifically, they are all members of a weekly gaming group where they play using human cooperation.
Five participants also stated that they had played \textit{Rocket League} in the past (\ParticipantThree{}, \ParticipantFive{}, \ParticipantSix{}, \ParticipantEight{}, \ParticipantTwelve{}).

%\revisionCHI{Finally, in the human cooperation session, the study supervisor acted as the copilot.
%The copilot is a PhD student in accessibility research, co-author of this paper, with a background in computer science and game design, and extensive experience playing video games multiple times per week.
%Their age range is 19--28 years, male, with moderate previous experience with \textit{Rocket League}.
%The copilot does not have any disabilities that affect gameplay.
%During the sessions, the copilot aimed to support participants in achieving their intended goals without imposing any decisions on them.}

Finally, in the human cooperation session, the study supervisor acted as the copilot. 
The copilot is a PhD student in accessibility research, co-author of this paper, with a background in computer science and game design, regular gaming experience, and moderate previous experience with \textit{Rocket League}. 
During the sessions, the copilot aimed to support participants in achieving their intended goals without imposing decisions on them.

\subsection{Data Analysis}
\label{ssec:metric_analysis}

\begin{figure}[htb]
    \centering
    \includegraphics[width=0.5\linewidth]{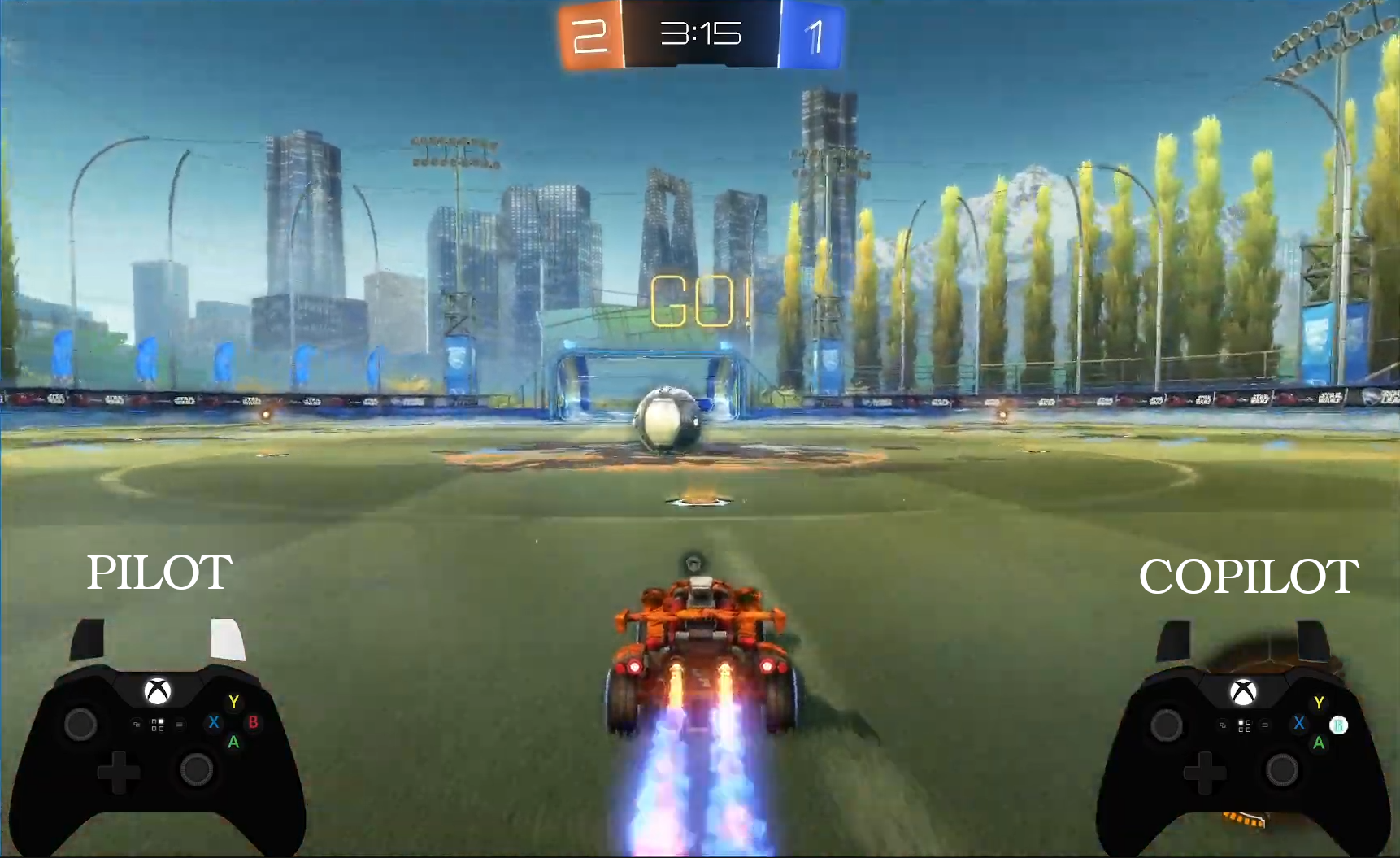}
    \caption{\revisionCHI{Screenshot of the augmented video used for the analysis of \ParticipantThree{}'s human cooperation session.}}
    \label{fig:video-overlay}
    \Description{Screenshot of a Rocket League match, to which an overlay is applied. The overlay depicts two xbox game controllers at the bottom corners of the image. The controller on the left is labeled "pilot"; the one on the right is labeled "copilot". The two controllers express with a different color the buttons that are currently being pressed/moved by the pilot and the copilot.
    In the example, the pilot is pressing the right trigger of the controller to accelerate, while the copilot is pressing the button B of the controller to boost.}
\end{figure}

% Data
\revisionCHI{During the gaming sessions, we collected ambient audio, screen recordings, and \textit{input command} logs.
For each gaming session, we merged these data sources into a single video by synchronizing audio and screen recordings and visually representing the activated \textit{input commands} using two on-screen game controllers, one for the pilot and one for the copilot.
For example, Fig.~\ref{fig:video-overlay} shows the pilot accelerating (right trigger is highlighted) and the copilot boosting  (B button is highlighted). 
These augmented videos allowed us to observe how players acted and coordinated their actions during gameplay, to analyze their verbal utterances, relevant behavioral cues (\textit{e.g.}, laughter, hesitation, silence), and conversations.
The \textit{game actions} subdivisions chosen by participants and the transcripts of the final interviews were also analyzed.}

% Analysis
\revisionCHI{Data were analyzed following the reflexive thematic analysis methodology~\cite{terry2017thematic}.
Conducting thematic analysis on observational data (\textit{e.g.}, video recordings) is an established practice in human-computer interaction and accessibility research~\cite{lemke2020motivators,schaadhardt2021understanding,taeb2024axnav,manzoni2025mapio}.}
\revisionCHI{The analysis combined both deductive and inductive aspects.
The deductive approach was applied to outline the first three themes.
The first two themes -- \textit{Effectiveness of the Support Provided by \system{}} (Section~\ref{ssec:configuration}) and \textit{User Experience} (Section~\ref{ssec:ux}) -- were based on the responses to the questions posed in the final interview that we designed based on previous literature (Section~\ref{ssec:protocol}).
Instead, the third theme -- \textit{Collaboration Confusion} (Section~\ref{ssec:confusion}) -- extends prior literature on automation confusion~\cite{cimolino2023automation}, previously defined solely for partial automation, to the broader context of shared control.
The inductive approach was instead used for the last theme -- \textit{Gameplay Interventions} (Section~\ref{ssec:gameplay_interventions}):
%and \textit{Communication Between Players} (Section~\ref{ssec:communication}).
%For these themes, w
we started with the initial codes extracted from the collected data, which were then coalesced into themes through subsequent discussions and refinements.}

\revisionCHI{The analysis of each participant's data began with the coders independently reviewing the two gaming session videos, producing an initial set of codes.
During this analysis, coders slowed down the video, analyzing it frame by frame when necessary.
Each code was associated with its timestamp to facilitate revision and discussion.
Successively, the final interview transcripts were analyzed, refining and expanding the initial codes
%, as participants’ reflections provided additional insights into their gaming sessions
.}

\revisionCHI{Two experiments were jointly analyzed by four researchers to collaboratively identify initial codes.
Subsequent experiments were randomly divided among the researchers.
Each experiment was coded by a main researcher and reviewed by a second researcher, who integrated the work with additional observations.}
\revisionCHI{Finally, all codes were collectively reviewed in dedicated meetings during which the research team discussed ambiguous cases, resolved disagreements, and revisited the corresponding video segments to ensure a shared interpretation.
The extracted codes were consolidated and organized into sub-themes and main themes through an iterative process of comparison among researchers.
The results of this analysis are reported in Section~\ref{sec:results-qualitative}.}

\subsubsection{Reflexivity Statement}

Reflexive thematic analysis acknowledges that researchers’ perspectives inevitably shape qualitative data interpretation~\cite{braun2019thematic}.
As such, it encourages researchers to reflect on their backgrounds and assumptions and how these may influence the analytic process.
In the following, we describe the perspectives we bring to this work.

All authors are researchers in the fields of human–computer interaction and accessibility.
The team includes both senior researchers -- with extensive experience working with people with disabilities -- and junior researchers.
Most authors have previously worked on video game accessibility, while one author was approaching this research area for the first time.

All authors regularly play video games on different platforms, including PC, console, and smartphone.
Some authors regularly use accessibility options during gameplay.
However, none of them had prior direct experience playing games using human cooperation or partial automation before the start of this project.
These experiences may have shaped how we interpreted participants’ accounts, for example, by influencing our understanding of common gameplay practices and expectations about the challenges faced by players with motor impairments.

%Given these different perspectives, the analysis was conducted collaboratively.
%The authors regularly discussed codes, interpretations, and emerging themes in order to reflect on their assumptions and reach shared interpretations of the data.

\section{Results}
\label{sec:results-qualitative}

In this section, we report the main results of the study.
We first describe the game configurations selected by the participants (Section~\ref{ssec:qualitative-selected-configuration}).
Then, we present the four main themes identified during the thematic analysis (Table~\ref{tab:themes})
%:
% \textit{Effectiveness of the Support Provided by \system{}} (Section~\ref{ssec:configuration}),
% \textit{User Experience} (Section~\ref{ssec:ux}),
% \textit{Collaboration Confusion} (Section~\ref{ssec:confusion}),
% \textit{Gameplay Interventions} (Section~\ref{ssec:gameplay_interventions}), and \textit{Communication Between Players} (Section~\ref{ssec:communication})
.
In many cases, we identified parallels in themes and topics across human cooperation and partial automation; thus, when appropriate, we analyze them comparatively.
%Thus, where appropriate, we present the results separately by the type of shared control support and analyze them comparatively.

\begin{table}[htb]
\caption{Themes and sub-themes identified through the reflexive thematic analysis}
\label{tab:themes}
\centering
\begin{tabular}{|l|l|}
\hline
\textbf{Theme} & \textbf{Sub-themes} \\ 
\hline
\multirow{2}{*}{\hyperref[ssec:configuration]{1. \textit{Effectiveness of the Support Provided by \system{}}}} 
    & \hyperref[sssec:configuration-usefulness-shared]{1. \textit{Usefulness of Shared Control}} (Section~\ref{sssec:configuration-usefulness-shared}) \\
    %& \hyperref[sssec:configuration-physical-setup]{2. \textit{Effectiveness of the Physical Setup}} \\
    & \hyperref[sssec:configuration-control-assignment]{2. \textit{Effectiveness of the Game Action Subdivision}} (Section~\ref{sssec:configuration-control-assignment}) \\
    \hline

\multirow{2}{*}{\hyperref[ssec:ux]{2. \textit{User Experience}}} 
    & \hyperref[sssec:ux-preferred_modality]{1. \textit{Preferred Support Modality}} (Section~\ref{sssec:ux-preferred_modality}) \\
    & \hyperref[sssec:ux-sentiment]{2. \textit{Participants' Sentiment}} (Section~\ref{sssec:ux-sentiment}) \\
    \hline

\multirow{2}{*}{\hyperref[ssec:confusion]{3. \textit{Collaboration Confusion}}} 
    & \hyperref[sssec:confusion-false]{1. \textit{Confusion Due to False Causation}} (Section~\ref{sssec:confusion-false}) \\
    & \hyperref[sssec:confusion-unexpected]{2. \textit{Confusion Due to Unexpected Actions}} (Section~\ref{sssec:confusion-unexpected}) \\
    \hline

\multirow{4}{*}{\hyperref[ssec:gameplay_interventions]{4. \textit{Gameplay Coordination}}}
    & \hyperref[sssec:gameplay_interventions-synergizing]{1. \textit{Effective Coordination by Synergizing}} (Section~\ref{sssec:gameplay_interventions-synergizing}) \\
    & \hyperref[sssec:gameplay_interventions-correction]{2. \textit{Effective Coordination by Correction}} (Section~\ref{sssec:gameplay_interventions-correction}) \\
    & \hyperref[sssec:gameplay_interventions-cooperation_ineffective]{3. \textit{Ineffective Coordination}} (Section~\ref{sssec:gameplay_interventions-cooperation_ineffective}) \\
    & \hyperref[sssec:gameplay_interventions-software-copilot-skills]{4. \textit{Effect of the Software Copilot's Skills}} (Section~\ref{sssec:gameplay_interventions-software-copilot-skills}) \\
\hline

%\multirow{6}{*}{\hyperref[ssec:communication]{5. \textit{Communication Between Players}}} 
%    & \hyperref[sssec:communication-instructing]{1. \textit{Instructing}} \\
%    & \hyperref[sssec:communication-signaling]{2. \textit{Signaling}} \\
%    & \hyperref[sssec:communication-planning]{3. \textit{Planning}} \\
%    & \hyperref[sssec:communication-commanding]{4. \textit{Commanding}} \\
%    & \hyperref[sssec:communication-motivating]{5. \textit{Motivating}} \\
%    & \hyperref[sssec:communication-small-talk]{6. \textit{Small Talk}} \\
%\hline
\end{tabular}
\Description{A table presenting the themes and sub-themes identified through reflexive thematic analysis. 
It is organized in two columns: the left column lists the main themes, 
while the right column specifies the corresponding sub-themes associated with each theme.}
\end{table}

\subsection{Selected \textit{Game Actions} Subdivisions}
\label{ssec:qualitative-selected-configuration}

Table~\ref{tab:configurations} shows how participants divided the \textit{game actions} with the copilot during the study.
\revisionCHI{Three participants (\ParticipantTwo{}, \ParticipantFour{}, \ParticipantSix{}) shared some \textit{game actions} with the copilot, while the others assigned distinct sets of \textit{game actions} to themselves and to the copilot.}
Two participants (\ParticipantSeven{} and \ParticipantTen{}) controlled a single \textit{game action}, while most participants controlled two (\ParticipantNine{}, \ParticipantFifteen{}, \ParticipantSixteen{}) or three (\ParticipantThree{}, \ParticipantFour{}, \ParticipantFive{}, \ParticipantEleven{}, \ParticipantTwelve{}).
Three participants controlled four or more (\ParticipantTwo{}, \ParticipantSix{}, \ParticipantEight{}).
Most participants took control of steering, whereas \ParticipantFifteen{} and \ParticipantSixteen{} chose not to.
Accelerating was controlled by nine participants, boosting by six, braking by five, jumping by four, and handbraking by two.

\begin{table}[t]
\centering
\caption{Participants' subdivisions of \textit{game actions} between pilot and copilot. 
P: pilot controlled the game action, C: copilot controlled the game action, O: overlapping (both the pilot and copilot controlled the game action)}
\label{tab:configurations}
\begin{tabular}{|l|c|c|c|c|c|c|c|c|c|c|c|c|c|}
\hline
\textbf{Action} &
  \textbf{\ParticipantTwo{}} &
  \textbf{\ParticipantThree{}} &
  \textbf{\ParticipantFour{}} &
  \textbf{\ParticipantFive{}} &
  \textbf{\ParticipantSix{}} &
  \textbf{\ParticipantSeven{}} &
  \textbf{\ParticipantEight{}} &
  \textbf{\ParticipantNine{}} &
  \textbf{\ParticipantTen{}} &
  \textbf{\ParticipantEleven{}} &
  \textbf{\ParticipantTwelve{}} &
  \textbf{\ParticipantFifteen{}} &
  \textbf{\ParticipantSixteen{}} \\ 
\hline
\textbf{Steering}     & \cellcolor{pilot}P & \cellcolor{pilot}P & \cellcolor{overlap}O & \cellcolor{pilot}P & \cellcolor{pilot}P & \cellcolor{pilot}P & \cellcolor{pilot}P & \cellcolor{pilot}P & \cellcolor{pilot}P & \cellcolor{pilot}P & \cellcolor{pilot}P & \cellcolor{copilot}C & \cellcolor{copilot}C \\ \hline
\textbf{Accelerating} & \cellcolor{pilot}P & \cellcolor{pilot}P & \cellcolor{pilot}P & \cellcolor{pilot}P & \cellcolor{pilot}P & \cellcolor{copilot}C & \cellcolor{pilot}P & \cellcolor{pilot}P & \cellcolor{copilot}C & \cellcolor{pilot}P & \cellcolor{copilot}C & \cellcolor{copilot}C & \cellcolor{pilot}P \\ \hline
\textbf{Braking}      & \cellcolor{copilot}C & \cellcolor{pilot}P & \cellcolor{pilot}P & \cellcolor{copilot}C & \cellcolor{pilot}P & \cellcolor{copilot}C & \cellcolor{pilot}P & \cellcolor{copilot}C & \cellcolor{copilot}C & \cellcolor{pilot}P & \cellcolor{copilot}C & \cellcolor{copilot}C & \cellcolor{copilot}C \\ \hline
\textbf{Jumping}         & \cellcolor{overlap}O & \cellcolor{copilot}C & \cellcolor{copilot}C & \cellcolor{copilot}C & \cellcolor{overlap}O & \cellcolor{copilot}C & \cellcolor{copilot}C & \cellcolor{copilot}C & \cellcolor{copilot}C & \cellcolor{copilot}C & \cellcolor{pilot}P & \cellcolor{pilot}P & \cellcolor{copilot}C \\ \hline
\textbf{Boosting}        & \cellcolor{pilot}P & \cellcolor{copilot}C & \cellcolor{copilot}C & \cellcolor{pilot}P & \cellcolor{copilot}C & \cellcolor{copilot}C & \cellcolor{pilot}P & \cellcolor{copilot}C & \cellcolor{copilot}C & \cellcolor{copilot}C & \cellcolor{pilot}P & \cellcolor{pilot}P & \cellcolor{pilot}P \\ \hline
\textbf{Handbraking}    & \cellcolor{copilot}C & \cellcolor{copilot}C & \cellcolor{copilot}C & \cellcolor{copilot}C & \cellcolor{pilot}P & \cellcolor{copilot}C & \cellcolor{pilot}P & \cellcolor{copilot}C & \cellcolor{copilot}C & \cellcolor{copilot}C & \cellcolor{copilot}C & \cellcolor{copilot}C & \cellcolor{copilot}C \\ \hline
\end{tabular}%
\Description{A table showing the game action subdivision chosen by participants for the experiment. Each row represents one of the Rocket League game actions. Each column represents a participant. A table cell contains the letter P if the action is assigned to the pilot, C if it's assigned to the copilot, and O if the control is overlapping, meaning both pilot and copilot can control it.
}
\end{table}

\begin{figure}[hbt]
    \centering
    \includegraphics[width=0.45\linewidth]{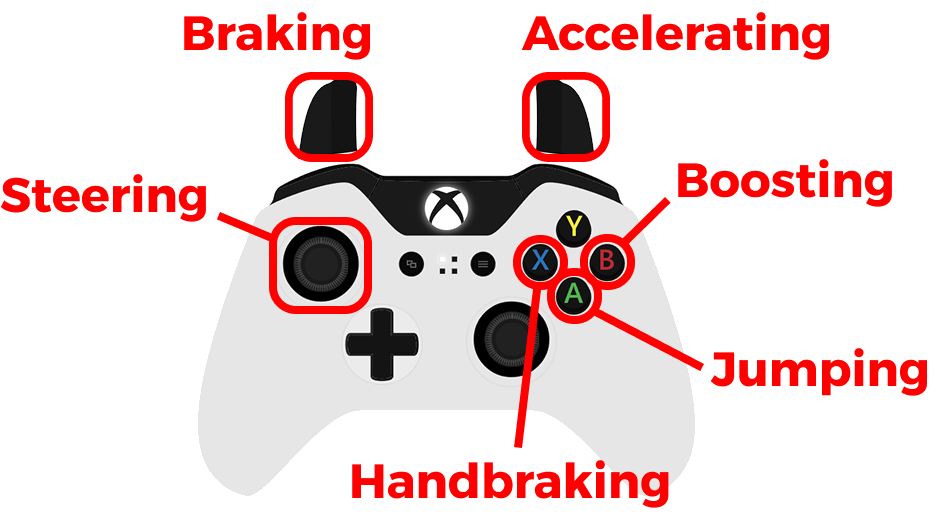}
    \caption{Default \textit{Rocket League} mapping of \textit{input commands} to \textit{game actions}}
    \label{fig:default-mapping}
    \Description{A drawing of an Xbox standard controller, on which the Rocket League default command mapping is drawn. Steering is controlled using the left analog stick; braking is controlled using the left trigger; acceleration is controlled using the right trigger; the buttons X, A, and B map respectively onto handbraking, jumping, and boosting.}
\end{figure}

%Considering the choice of the controller, o
Only \ParticipantTen{} used the \textit{Xbox Adaptive Controller}~\cite{xboxAdaptiveController}, along with two external buttons to control steering (one button to steer in each direction).
All other participants used a standard \textit{Xbox} controller.
The majority of participants adopted the game default mapping of \textit{input commands} to \textit{game actions} (Fig.~\ref{fig:default-mapping}): steering was controlled via the left analog stick, accelerating through the right trigger, braking through the left trigger, jumping via the \textit{A} button, boosting via the \textit{B} button, and handbraking via the \textit{X} button.
Some participants implemented custom modifications: \ParticipantFour{} remapped accelerator and brake to the \textit{B} and \textit{X} buttons, respectively; \ParticipantEight{} assigned handbrake to the \textit{X} button and boost to the \textit{A} button; both \ParticipantNine{} and \ParticipantSixteen{} mapped the accelerator to the \textit{A} button.

The analysis of the gameplay sessions revealed that some participants did not actively use all the \textit{game actions} they chose to control.
In human cooperation, \ParticipantThree{}, \ParticipantFour{}, \ParticipantSix{}, and \ParticipantEleven{} never used the brake, while \ParticipantSix{} used the handbrake only once.
Similarly, in partial automation, \ParticipantFour{} did not use the brake, and \ParticipantSix{} and \ParticipantEight{} used the handbrake once and twice, respectively.
\ParticipantTwelve{} activated the boost only once, later acknowledging that they forgot about it.
%that it was available.

\subsubsection{\revisionCHI{Relationship Between Participant Characteristics and Selected Configurations}}
\label{sssec:configuration-vs-characteristics}

\revisionCHI{By relating the configurations selected by the participants (Table~\ref{tab:configurations}) and their characteristics (Table~\ref{tab:demographic-data}) we identified several trends.}
\revisionCHI{Participants who perceived more gaming difficulties often assigned a larger portion of \textit{game actions} to the copilot.
For example, \ParticipantSeven{} and \ParticipantTen{} both controlled only one \textit{game action}, delegating the majority of controls to the copilot.
Similarly, \ParticipantNine{} controlled only two actions.
All three reported \quots{quite a bit} of difficulty (4 out of 5) in playing video games.
In contrast, \ParticipantThree{} and \ParticipantFive{}, who reported only a \quots{little} (2 out of 5) and \quots{moderate} (3 out of 5) gameplay difficulty, controlled three \textit{game actions} each.
Instead, \ParticipantTwo{}, despite reporting \quots{quite a bit} of gameplay difficulty, chose to control four \textit{game actions}.}

\revisionCHI{Console/PC gamers (\ParticipantTwo{}, \ParticipantThree{}, \ParticipantFive{}, \ParticipantSix{}, \ParticipantEight{}, \ParticipantEleven{}, \ParticipantTwelve{}) generally kept three or more \textit{game actions} under their control.
This suggests that familiarity with traditional gamepads may have lowered the perceived effort for managing multiple \textit{game actions}.
In contrast, participants who primarily play on smartphones or tablets (\ParticipantFour{}, \ParticipantSeven{}, \ParticipantNine{}, \ParticipantTen{}, \ParticipantFifteen{}, \ParticipantSixteen{}), who may have less experience with multi-input controllers, typically selected at most two \textit{game actions}.
Only \ParticipantFour{} controlled three actions, but one of them was shared with the copilot.}
\revisionCHI{Participants with previous experience with \textit{Rocket League} (\ParticipantThree{}, \ParticipantFive{}, \ParticipantSix{}, \ParticipantEight{}, \ParticipantFifteen{}) always selected at least three \textit{game actions} each.
Notably, all five maintained direct control over both steering and acceleration.}

\subsection{Effectiveness of the Support Provided by \system{}}
\label{ssec:configuration}

We identified two topics related to the effectiveness of the provided support: usefulness of shared control (Section~\ref{sssec:configuration-usefulness-shared}),
%effectiveness of the physical setup (Section~\ref{sssec:configuration-physical-setup}),
and effectiveness of the \textit{game actions} subdivision (Section~\ref{sssec:configuration-control-assignment}).

\subsubsection{Usefulness of Shared Control}
\label{sssec:configuration-usefulness-shared}
Shared control allowed all participants to play, thanks to the fact that it reduced the number of \textit{input commands} that they had to control.
\revisionCHI{Indeed, most participants (\ParticipantFour{}, \ParticipantSeven{}, \ParticipantNine{}, \ParticipantTen{}, \ParticipantTwelve{}, \ParticipantFifteen{}, \ParticipantSixteen{}) reported that they would not have been able to play without such support.
%of a copilot.
In particular, \ParticipantTwelve{} specified that they tried playing \textit{Rocket League} in the past without success.
Conversely, all others participants indicated they could have played without support, but also acknowledged that this would result in increased difficulty or reduced performance.
In particular, \ParticipantTwo{} and \ParticipantThree{} reported that they would not have been able to play at the level they desired:}
\revisionCHI{
\begin{displayquote}
\ParticipantTwo{}: \it
[I would have been able to play on my own], though it probably would’ve been harder at first and maybe not in the way I like to play. After a while, I think it might become a bit frustrating.\footnote{All questions and quotes reported in the following are translated from \anon[the language spoken by the participants]{Italian}.}
\end{displayquote}
\ParticipantSix{} noted that they can currently play without support, but possibly not in the future due to their degenerative condition, while \ParticipantEleven{} stated that matching the performance achieved in the experiment would require practice.
Finally, shared control also reduced the perceived cognitive load.
For example, \ParticipantSix{} highlighted that receiving assistance for boosting relieved them from thinking about when to use it and from worrying about the boost fuel left in the car.}

\pa{}
Aside from the benefits common to both shared control approaches, partial automation also increases the autonomy with respect to human cooperation by removing the need for a human copilot (\ParticipantThree{}, \ParticipantSeven{}, \ParticipantEight{}):
\revisionCHI{
\begin{displayquote}
\ParticipantEight{}: \it
I was autonomous; I did not have to ask somebody else for help when I wanted to do something.
\end{displayquote}
Indeed, all participants, except for \ParticipantSix{}, \ParticipantTen{}, and \ParticipantEleven{}, reported that they would continue to use partial automation at home, also with other games (\ParticipantThree{}, \ParticipantFour{}, \ParticipantFive{}, \ParticipantTwelve{}, \ParticipantFifteen{}).}
\ParticipantTen{}, instead, acknowledged its potential usefulness, but considered it irrelevant for the games they usually play (\textit{e.g.}, card games, chess).

%\subsubsection{Effectiveness of the Physical Setup}
%\label{sssec:configuration-physical-setup}

%\revisionCHI{Despite having selected their preferred controller and \textit{input commands}, some participants experienced difficulties physically interacting with the controls (\ParticipantFour{}, \ParticipantFive{}, \ParticipantEight{}, \ParticipantNine{}, \ParticipantTen{}, \ParticipantTwelve{}). 
%For example, \ParticipantEight{} struggled to press the \textit{Y} button to brake, which is something they did not realize while testing the configuration.}
%
%Some difficulties were due to the provided hardware.
%\ParticipantFive{} and \ParticipantEight{} usually play with a \textit{PlayStation} controller~\cite{ps5Controller} and did not feel as comfortable using the \textit{Xbox} controller~\cite{xboxController}.
%Similarly, \ParticipantTen{} had difficulty using two buttons from the \textit{Logitech Adaptive Gaming Kit} to control the steering, and sometimes pressed both by mistake.
%The participant later explained that they would have preferred a joystick like those on wheelchairs, as they were already accustomed to them.

\subsubsection{Effectiveness of the \textit{Game Actions} Subdivision}
\label{sssec:configuration-control-assignment}

\revisionCHI{Most participants (\ParticipantThree{}, \ParticipantFour{}, \ParticipantFive{}, \ParticipantTen{}, \ParticipantTwelve{}, \ParticipantFifteen{}, \ParticipantSixteen{}) were satisfied with the \textit{game actions} subdivision between pilot and copilot (Section~\ref{ssec:qualitative-selected-configuration}).}
Instead, others reported that they would change something if they kept playing.
\ParticipantTwo{} would have preferred to control the jump action personally, due to a perceived ineffectiveness in the way the copilot was using it (Section~\ref{sssec:gameplay_interventions-cooperation_ineffective}).
\revisionCHI{\ParticipantSix{}, instead, explained that they had underestimated the importance of boosting for the gameplay, and that they would have assigned it to themselves.}
\revisionCHI{Finally, \ParticipantSeven{} and \ParticipantEleven{}
pointed out that being able to control all \textit{game actions} would have been more satisfying:}
\revisionCHI{
\begin{displayquote}
\ParticipantSeven{}: \it
One would like to do everything by themselves, but it is probably not realistic. If I were to keep playing, I would try to play with no support, but I am not sure I would be able to do it.
\end{displayquote}
}

\subsection{User Experience}
\label{ssec:ux}
This section analyzes participants' preferred support modality between human cooperation and partial automation (Section~\ref{sssec:ux-preferred_modality}) and their sentiment (Section~\ref{sssec:ux-sentiment}).

\subsubsection{Preferred Support Modality}
\label{sssec:ux-preferred_modality}

\hc{}
Seven participants (\ParticipantFour{}, \ParticipantSix{}, \ParticipantNine{}, \ParticipantEleven{}, \ParticipantTwelve{}, \ParticipantFifteen{}, \ParticipantSixteen{}) favored human cooperation over partial automation.
Many participants motivated this preference by reporting that
%A relevant aspect motivating this preference is that, as reported by many participants,
cooperating with another human being feels more \quots{natural} (\ParticipantFour{}, \ParticipantSix{}, \ParticipantSeven{}, \ParticipantEight{}, \ParticipantNine{}, \ParticipantEleven{}, \ParticipantFifteen{}).
\revisionCHI{For example,
\ParticipantNine{} could not understand the logic behind the software copilots' jumps, which made coordination difficult (Section~\ref{sssec:gameplay_interventions-cooperation_ineffective}).}
In addition, the interaction with a human copilot was perceived as more natural due to the possibility to communicate (\ParticipantFour{}, \ParticipantFive{}, \ParticipantSix{}, \ParticipantSeven{}, \ParticipantNine{}, \ParticipantEleven{}, \ParticipantTwelve{}, \ParticipantFifteen{}), which, instead, was not possible in partial automation.
%(Section~\ref{ssec:communication}).
\ParticipantSix{} and \ParticipantFifteen{} noted that this allowed them to share their strategy with the copilot, or to tell them which \textit{game actions} to execute.
% For example:}
% %
% \begin{displayquote}
% \revisionCHI{
% \ParticipantSix{}: \it
% Ok, now let's push [the ball] to the other side, so that [the opponent] wastes more time.
% }
% \end{displayquote}
Finally, \ParticipantFour{} preferred human cooperation because it was perceived to correct their steering errors more effectively.

\pa{}
\revisionCHI{Other participants favored partial automation (\ParticipantTwo{}, \ParticipantThree{}, \ParticipantFive{}, \ParticipantSeven{}, \ParticipantEight{}), reporting several motivations.}
First, \ParticipantThree{}, \ParticipantSeven{}, and \ParticipantEight{} observed that partial automation increases autonomy, as it allows playing without the need for another person (Section~\ref{sssec:configuration-usefulness-shared}).
\revisionCHI{Second, for \ParticipantTwo{} and \ParticipantThree{} the software copilot played better than the human copilot, it was more predictable, and therefore it allowed them to focus more on the game and increase their performance (Section~\ref{ssec:gameplay_interventions}):}
\begin{displayquote}
\revisionCHI{
\ParticipantThree{}: \it
I did not like how the [human] copilot was supporting me. With the [software copilot] it was much easier: I knew what I had to do and what it had to do.
}
\end{displayquote}
Third, \ParticipantTwo{}, \ParticipantThree{}, and \ParticipantSeven{} perceived a higher level of support in partial automation, despite the \textit{game actions} subdivision being actually the same.
For example, although \ParticipantTwo{} shared jump control with the copilot in both sessions, they relied more heavily on the copilot in partial automation, using the jump action much less frequently themselves (4 vs. 22 times), while the software copilot used it much more than the human copilot (72 vs. 13 times).

\subsubsection{Participants' Sentiment}
\label{sssec:ux-sentiment}

Participants generally expressed enjoyment and engagement while playing in both human cooperation and partial automation.
However, some also expressed dissatisfaction with the game (\ParticipantFour{}, \ParticipantSeven{}, \ParticipantNine{}, \ParticipantTen{}, \ParticipantFifteen{}).
In particular, \ParticipantFour{} and \ParticipantSeven{} reported that they did not like driving games and therefore did not fully enjoy the experience.
Others expressed frustration while gaming.
For instance, \ParticipantNine{}, \ParticipantTen{}, and \ParticipantFifteen{} were dissatisfied with their own skill level and blamed themselves when something did not go as planned.
\revisionCHI{Notably, \ParticipantFifteen{} attributed all the credit to the copilot, whether human or software, when informed that they were winning.}

\hc{}
Most participants showed a strong sense of engagement in the game. For example, \ParticipantFive{}, \ParticipantSix{}, \ParticipantSeven{}, \ParticipantNine{} reacted with excitement to certain game situations, such as a missed goal, and \ParticipantFive{} congratulated themselves \quots{Good job!}.
Even those who reported not enjoying the game still expressed joy when scoring a goal or making a good play (\ParticipantSeven{}, \ParticipantNine{}, \ParticipantTen{}). 
\ParticipantSix{} commented on plays, developing a sense of complicity with the copilot \quots{Well done, keep boosting like this!}.
Although preferring partial automation, \ParticipantEight{} particularly appreciated the fun of playing with another person. 

\pa{}
Even in partial automation, participants felt engaged with the game.
For example, \ParticipantSix{} celebrated when scoring a goal and reported feeling in control and involved, despite some \textit{game actions} being delegated to the software copilot.
\begin{displayquote}
\revisionCHI{
\ParticipantSix{}: \it
When things were working well, I felt part of the action because I was doing everything else.
}
\end{displayquote}
\ParticipantSeven{}, \ParticipantEleven{}, and \ParticipantFifteen{} encouraged themselves during key actions (\textit{e.g.}, \quots{I got this, I can score!}) and \ParticipantSeven{} congratulated themselves when doing well.
\revisionCHI{Some participants also expressed excitement specifically about the partial automation system.
For example, \ParticipantFifteen{} commented: \quots{I'm happy I got to try the system!}.}

\subsection{Collaboration Confusion}
\label{ssec:confusion}

Cimolino et al.~\cite{cimolino2023automation} previously noted that partial automation \quots{can also make players confused about how the game is controlled}, a phenomenon they label \textit{automation confusion}.
We also observed a few situations in which the participants appeared confused, both in human cooperation and in partial automation sessions. 
In this section, we identify two sources of confusion and analyze how they impacted players' behavior.
We argue that the theory of automation confusion, originally formulated for partial automation only~\cite{cimolino2023automation}, can be extended to the general case of shared control, thus also including human cooperation.
Therefore, we refer to the phenomenon more generally as \textit{collaboration confusion}.

\subsubsection{Confusion Due to \textit{False Causation}~\cite{cimolino2023automation}}
\label{sssec:confusion-false}

Participants got confused when they did not remember which \textit{game actions} were assigned to them and which to the copilot (\ParticipantThree{}, \ParticipantFive{}, \ParticipantSeven{}, \ParticipantTen{}).
For example, \ParticipantTen{} could not understand why their car was frequently jumping, not remembering that the game action was under the copilot's control.
Confusion was also caused by attributing the observed actions to the wrong player (\ParticipantSix{}, \ParticipantFifteen{}).
For example, \ParticipantSix{}, who shared jump control with the copilot, believed they had performed an effective jump, while in fact it had been executed by the software copilot.
\revisionCHI{Conversely, \ParticipantFifteen{} in some cases blamed themselves, when in fact the copilot had control of the relevant action.}
For example, \ParticipantFifteen{} correctly boosted to reach the ball, but the software copilot failed to hit it.

\subsubsection{Confusion Due to Unexpected Actions}
\label{sssec:confusion-unexpected}
\begin{figure}[htb]
    \centering
    \begin{subfigure}[t]{0.32\linewidth}
        \includegraphics[width=\linewidth]{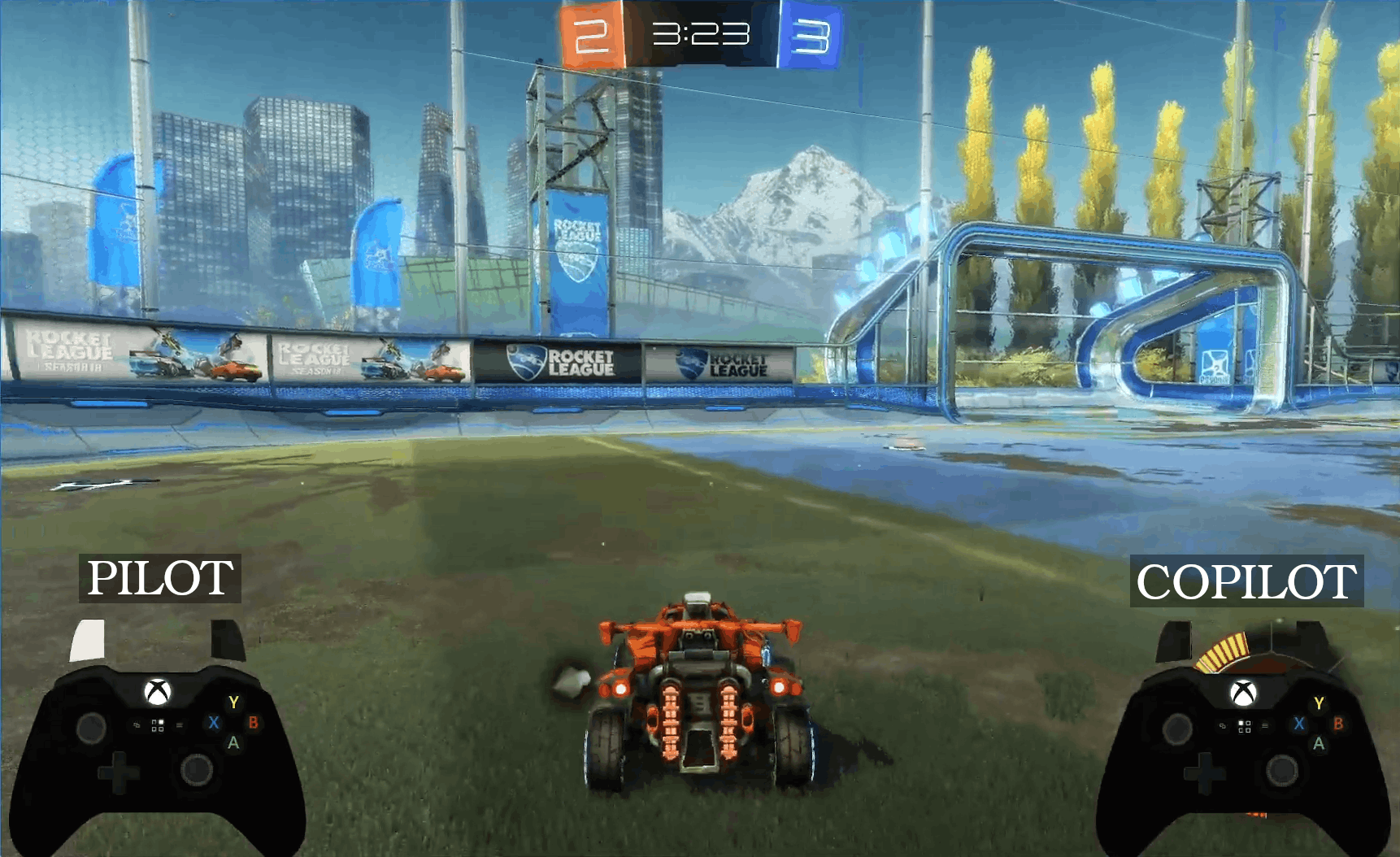}
        \caption{\revisionCHI{\ParticipantEleven{} is going in reverse since the ball is behind the car.}}
        \label{fig:gameplay-confusion-1}
        \Description{Screenshot of the augmented video used for the analysis of P10's partial automation session. P10 and their copilot are controlling an orange car. The car is currently going backwards.
        The pilot's controller shows the left trigger (the button for braking/reversing) being pressed.
        The copilot's controller shows no input.}
    \end{subfigure}
    \hfill
    \begin{subfigure}[t]{0.32\linewidth}
        \includegraphics[width=\linewidth]{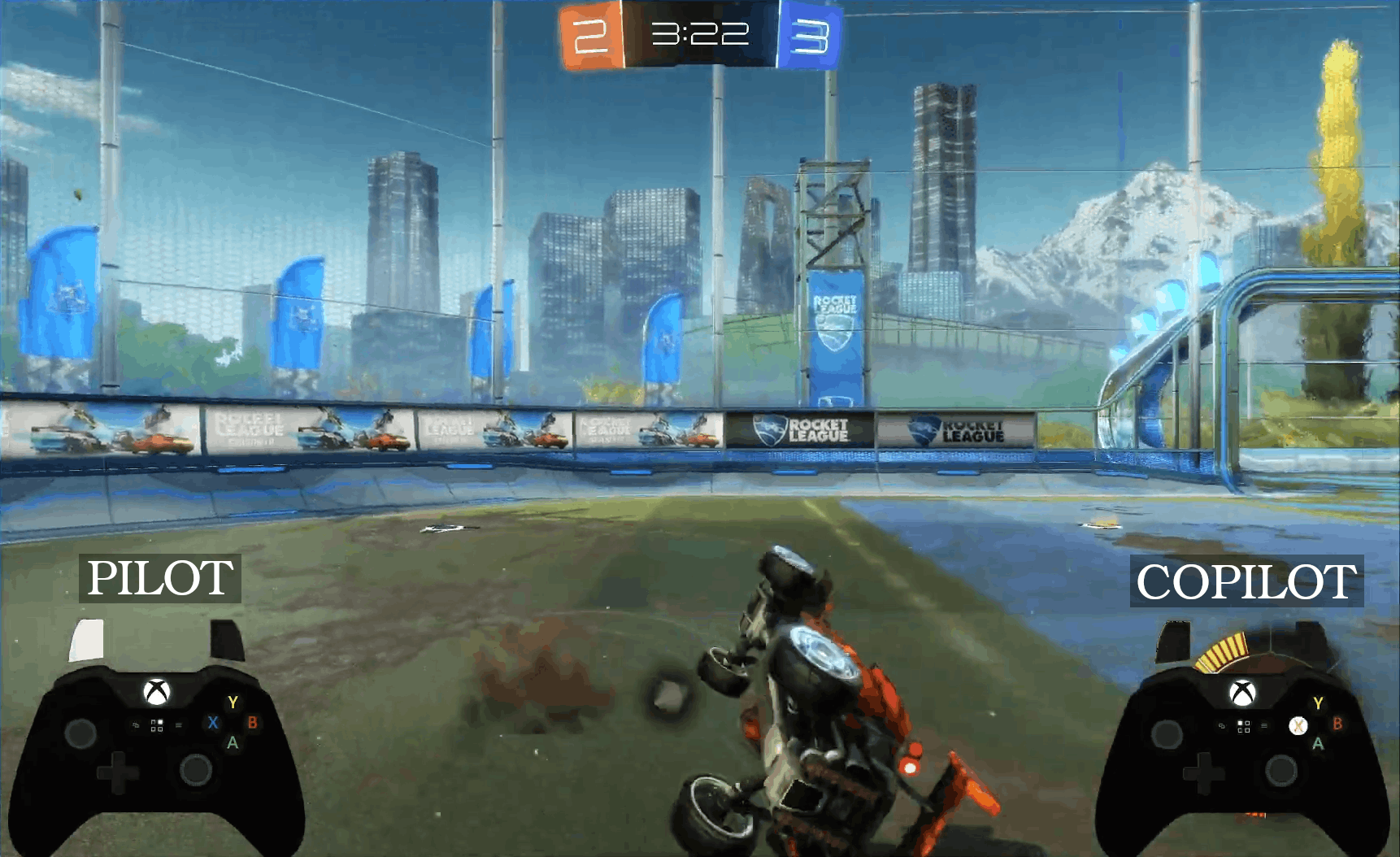}
        \caption{\revisionCHI{The copilot jumps, as turning around is easier while the car is in the air.}}
        \label{fig:gameplay-confusion-2}
        \Description{Screenshot of the augmented video used for the analysis of P10's partial automation session. P10 and their copilot are controlling an orange car. The car is currently in the air, upside down, jumping backwards and right.
        The pilot's controller shows the left trigger (braking/reversing) being pressed.
        The copilot's controller shows the button X (for handbrake) being pressed.}
    \end{subfigure}
    \hfill
    \begin{subfigure}[t]{0.32\linewidth}
        \includegraphics[width=\linewidth]{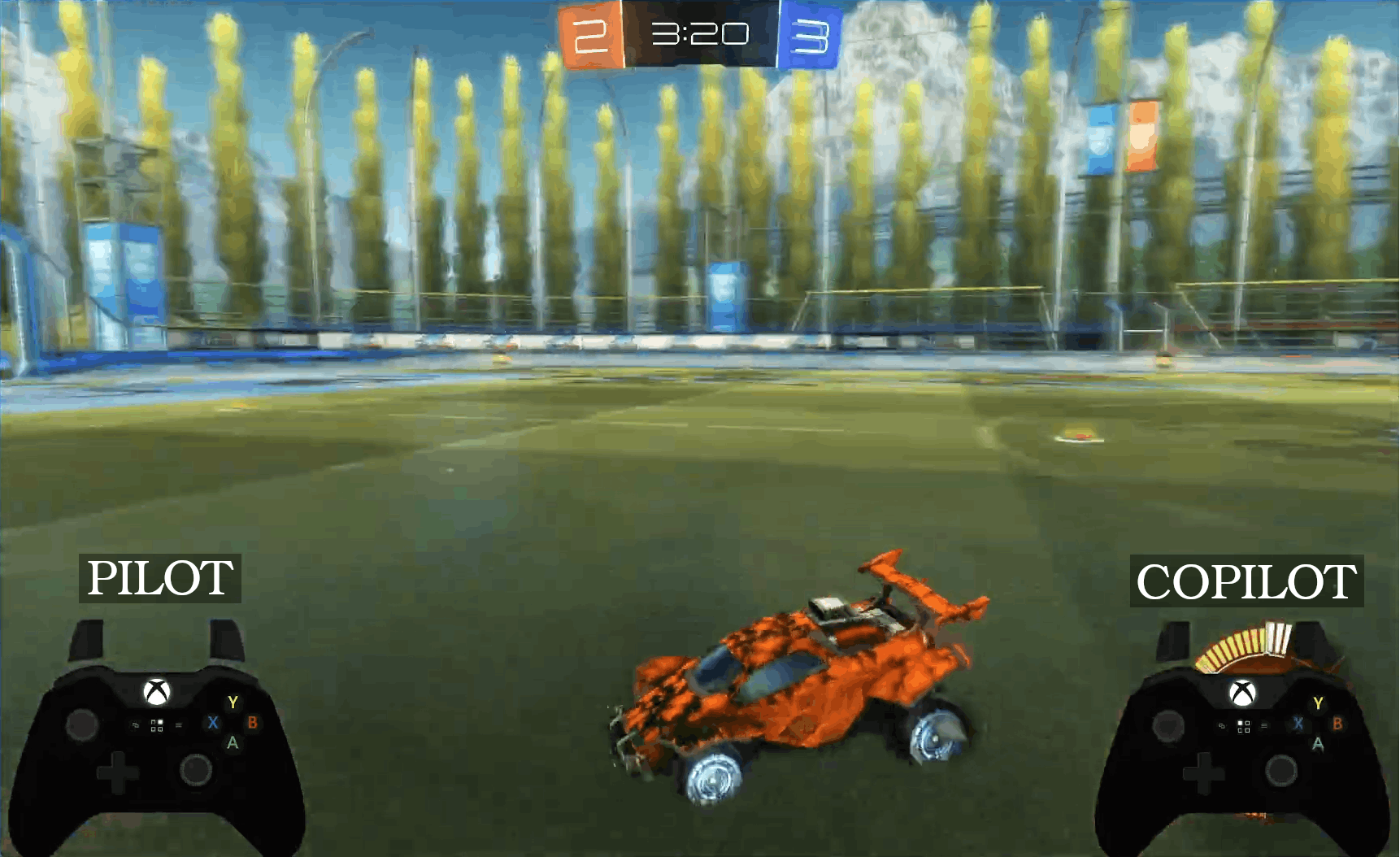}
        \caption{\revisionCHI{\ParticipantEleven{} does not understand the copilot's intentions and lands in a wrong direction.}}
        \label{fig:gameplay-confusion-3}
        \Description{Screenshot of the augmented video used for the analysis of P10's partial automation session. P10 and their copilot are controlling an orange car. The car is currently not moving, facing left.
        The pilot's controller shows no input.
        The copilot's controller shows no input either.}
    \end{subfigure}
    \caption{\revisionCHI{An example of \textit{Confusion due to unexpected actions} in \ParticipantEleven{}'s partial automation gameplay.}}
    \label{fig:gameplay-confusion}
\end{figure}

The second form of confusion was caused by copilot's actions whose intent was not clear to the pilot.
This problem was particularly common in partial automation, since the software copilot adopts advanced game tactics based on frequent jumps to move more quickly and to change direction rapidly (Section~\ref{sssec:gameplay_interventions-synergizing}).
An example is shown in Fig.~\ref{fig:gameplay-confusion}.
This kind of confusion can be mapped onto the \quots{No-Rule} mental model error defined by Cimolino et al.~\cite{cimolino2023automation}.
As a consequence, sometimes \ParticipantFour{} stopped playing completely for a while, a behavior that Cimolino et al. call \quots{Contemplation}.
To mitigate this issue, in the final interviews, participants proposed to use feedback cues to clarify the copilot's actions (\ParticipantSeven{}, \ParticipantNine{}, \ParticipantFifteen{}).
For example, \ParticipantNine{} proposed implementing on-screen indicators, such as directional arrows, to signal in advance when the copilot is about to intervene.
Instead, \ParticipantFifteen{} imagined receiving proactive feedback from the copilot in natural language.
The participant did not deem such feedback essential for gameplay, but expected it to significantly enhance comprehension of the copilot's intentions.

\subsection{Gameplay Coordination}
\label{ssec:gameplay_interventions}

We observed that players cooperated during gameplay through two possible approaches:
synergizing with the other player's actions (Section~\ref{sssec:gameplay_interventions-synergizing}) or correcting them (Section~\ref{sssec:gameplay_interventions-correction}).
The cooperation proved to be effective in many cases, but ineffective cooperation occasions also occurred (Section~\ref{sssec:gameplay_interventions-cooperation_ineffective}).
Finally, we discuss how playing with a high-performing agent as a software copilot affected coordination (Section~\ref{sssec:gameplay_interventions-software-copilot-skills}).

\subsubsection{Effective Coordination by Synergizing}
\label{sssec:gameplay_interventions-synergizing}

\hc{}
Pilot and copilot were able to effectively coordinate in several interactions.
For example, the copilot effectively supported the pilot's steering by using boost and handbrake (\ParticipantTwo{}, \ParticipantThree{}, \ParticipantFour{}, \ParticipantSix{}, \ParticipantSeven{}, \ParticipantNine{}, \ParticipantEleven{}, \ParticipantSixteen{}).
Similarly, the copilot frequently intervened at the right time with jumps that resulted in goals (\ParticipantSix{}, \ParticipantEleven{}, \ParticipantSixteen{}).
Cooperation appeared particularly effective in defensive situations, with good saves and recoveries (\ParticipantThree{}, \ParticipantNine{}, \ParticipantFifteen{}), as well as in some offensive moments, where coordination for shots and goals was fluid (\ParticipantEleven{}, \ParticipantFifteen{}, \ParticipantSixteen{}).

\begin{figure}[htb]
    \centering
    \begin{subfigure}[t]{0.32\linewidth}
        \includegraphics[width=\linewidth]{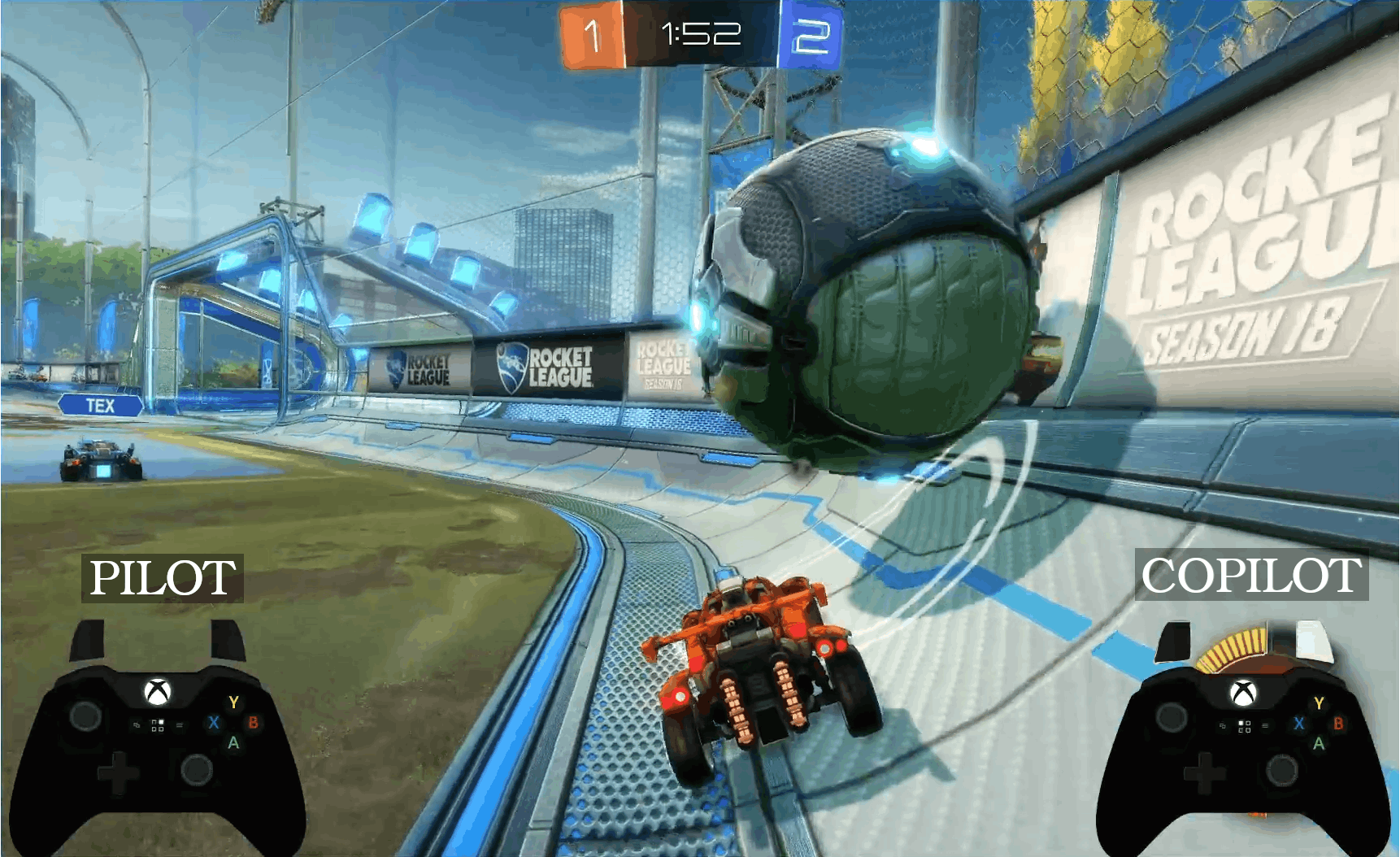}
        \caption{\revisionCHI{\ParticipantTwelve{} is heading towards the ball while the copilot is accelerating.}}
        \label{fig:gameplay-braking-1}
        \Description{Screenshot of the augmented video used for the analysis of P11's partial automation session. P11 and their copilot are controlling an orange car. The car is driving on the edges of the arena, moving forward towards the ball.
        The pilot's controller shows no inputs.
        The copilot's controller shows the right trigger (acceleration) being pressed.}
    \end{subfigure}
    \hfill
    \begin{subfigure}[t]{0.32\linewidth}
        \includegraphics[width=\linewidth]{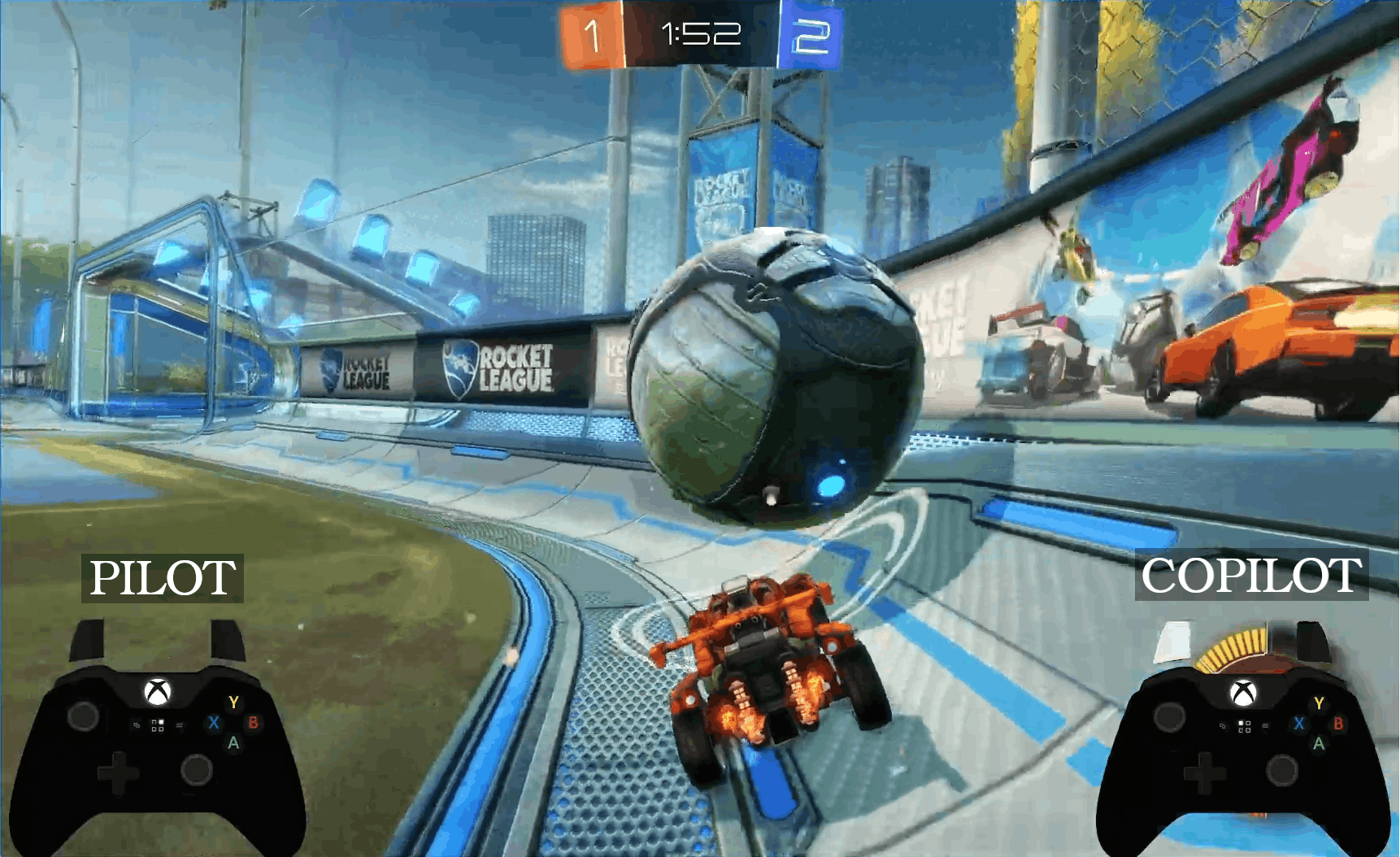}
        \caption{\revisionCHI{The copilot brakes to avoid passing beyond the ball.}}
        \label{fig:gameplay-braking-2}
        \Description{Screenshot of the augmented video used for the analysis of P11's partial automation session. P11 and their copilot are controlling an orange car. The car is driving on the edges of the arena, with the ball close to them, slightly to their right.
        The pilot's controller shows no inputs.
        The copilot's controller shows the left trigger (braking) being pressed.}
    \end{subfigure}
    \hfill
    \begin{subfigure}[t]{0.32\linewidth}
        \includegraphics[width=\linewidth]{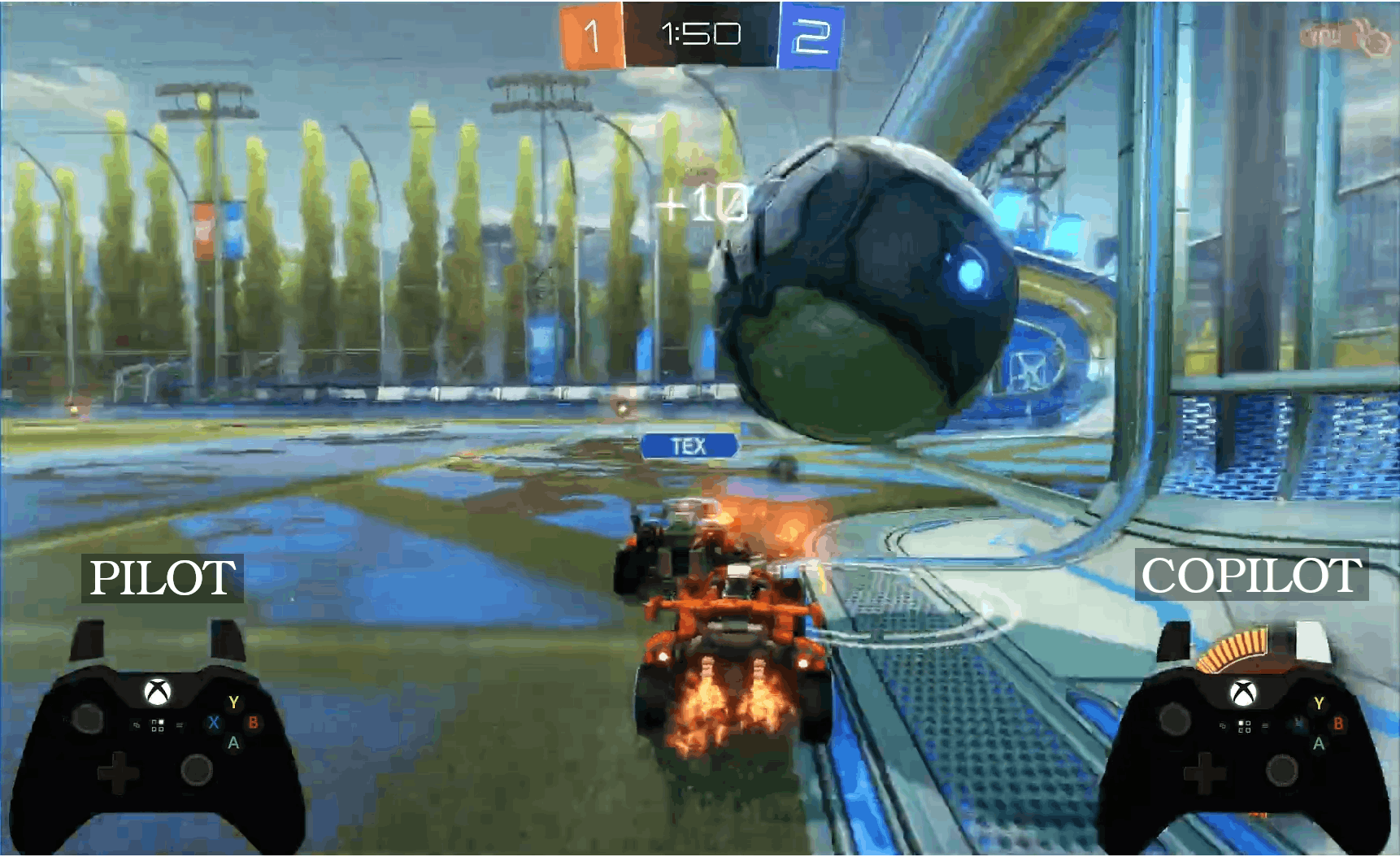}
        \caption{\revisionCHI{Once the ball rolls towards the goal, the car is in the right position to score.}}
        \label{fig:gameplay-braking-3}
        \Description{Screenshot of the augmented video used for the analysis of P11's partial automation session. P11 and their copilot are controlling an orange car. The car is in front of the opponent's goal. The ball is in the air between the car and the goal. The opponent is to the side of the ball, between the goal and the center of the field.
        The pilot's controller shows no inputs.
        The copilot's controller shows the right trigger (acceleration) being pressed.}
    \end{subfigure}
    \caption{\revisionCHI{An example of \textit{Effective Coordination by Synergizing} in \ParticipantTwelve{}'s partial automation gameplay.}}
    \label{fig:gameplay-braking}
\end{figure}

\pa{}
As in human cooperation, the software copilot could effectively coordinate with the pilot.
For example, the copilot occasionally slowed down, allowing the pilot to perform accurate hits on the ball (\ParticipantSeven{}, \ParticipantTen{}, \ParticipantTwelve{}), as shown in Fig.~\ref{fig:gameplay-braking}.
Furthermore, the copilot frequently performed well-timed interventions, for example, through jumps (\ParticipantSix{}) or boosts (\ParticipantThree{}, \ParticipantSix{}, \ParticipantSixteen{}).
Similarly to human cooperation, good synergy resulted in effective shots (\ParticipantTwo{}, \ParticipantThree{}, \ParticipantSeven{}) and saves (\ParticipantThree{}, \ParticipantFive{}, \ParticipantSeven{}, \ParticipantFifteen{}). 

\subsubsection{Effective Coordination by Correcting}
\label{sssec:gameplay_interventions-correction}
\revisionCHI{In some cases, precisely controlling \textit{game actions} was difficult for the pilot.
For example, \ParticipantFour{}, \ParticipantTen{}, and \ParticipantEleven{} struggled to steer accurately.
In such cases, the copilot sometimes intervened by correcting the pilot's actions using their own commands.}

\begin{figure}[htb]
    \centering
    \begin{subfigure}[t]{0.32\linewidth}
        \centering
        \includegraphics[width=\linewidth]{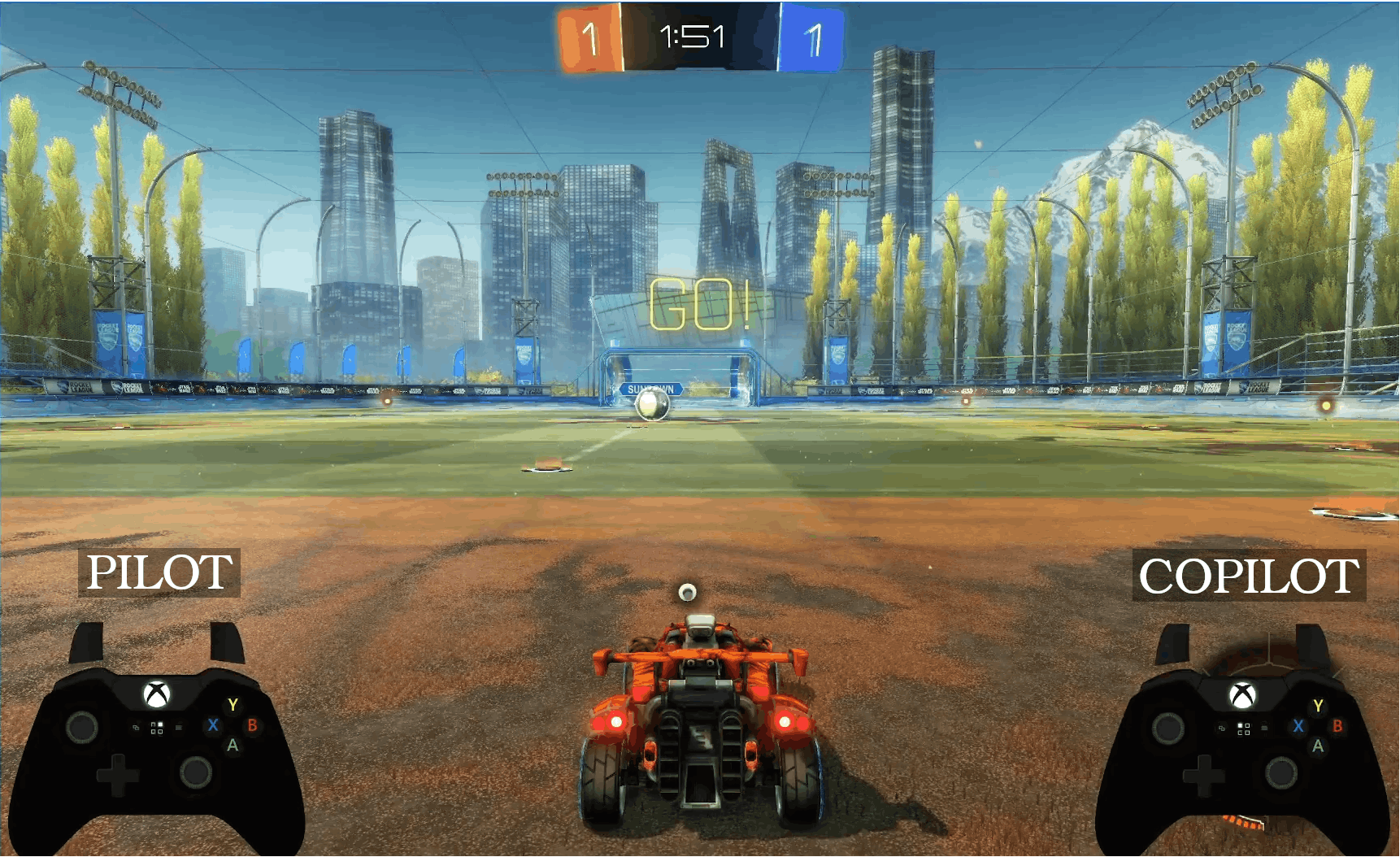}
        \caption{\revisionCHI{The game has kicked off, but \ParticipantFour{} is not accelerating.}}
        \label{fig:gameplay-correction-1}
        \Description{Screenshot of the augmented video used for the analysis of P3's human cooperation session. P3 and their copilot are controlling an orange car. The game has just kicked off: both cars are on their own side of the field, the ball is in the center.
        The pilot's controller shows no inputs.
        The copilot's controller shows no inputs either.}
    \end{subfigure}
    \hspace{0.02\linewidth}
    \begin{subfigure}[t]{0.32\linewidth}
        \centering
        \includegraphics[width=\linewidth]{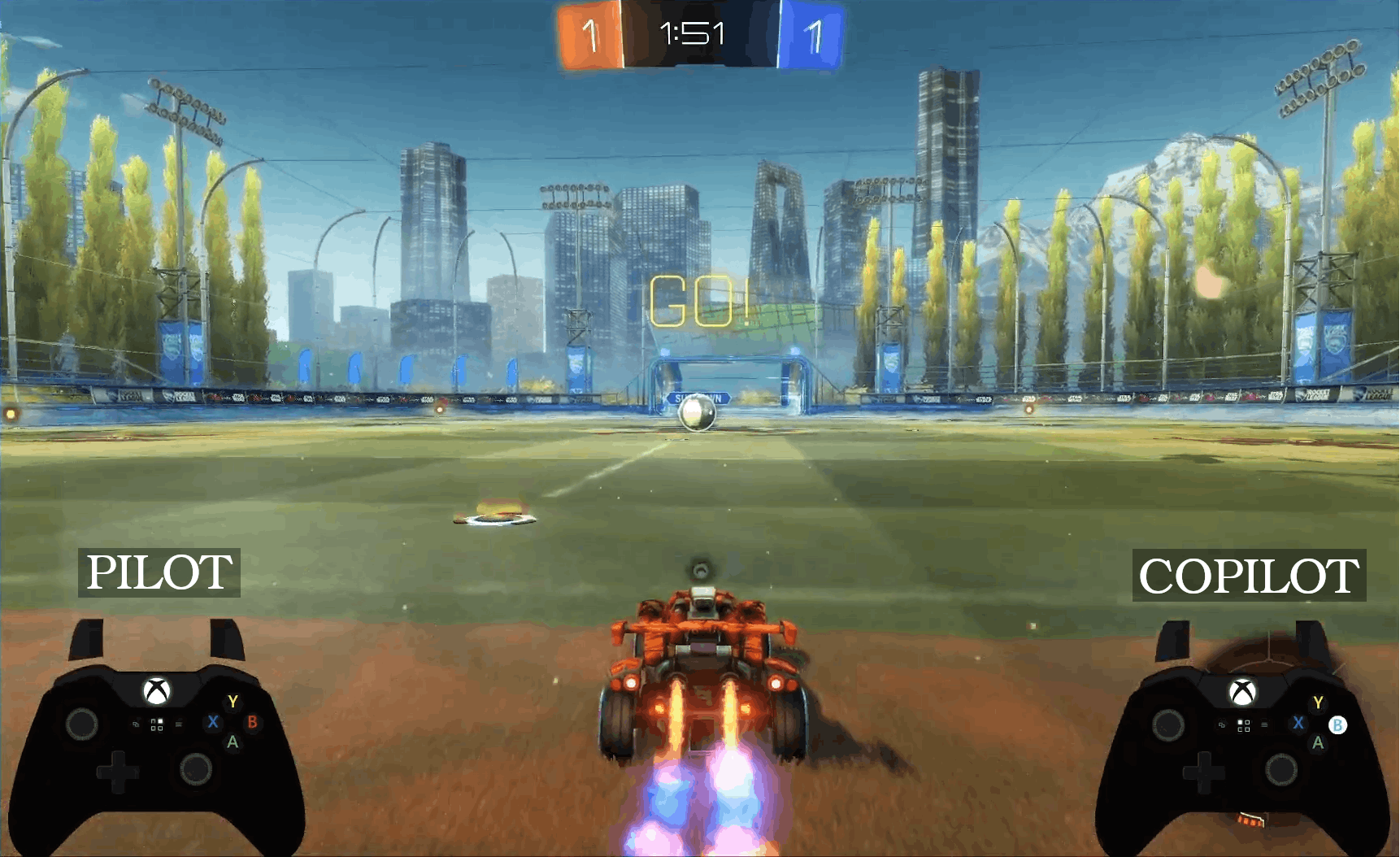}
        \caption{\revisionCHI{The copilot intervenes by boosting, forcing the car to start moving.}}
        \label{fig:gameplay-correction-2}
        \Description{Screenshot of the augmented video used for the analysis of P3's human cooperation session. P3 and their copilot are controlling an orange car. The game has just kicked off: the car is boosting towards the center of the field, where the ball is standing.
        The pilot's controller shows no inputs.
        The copilot's controller shows the B button (boost) being pressed.}
    \end{subfigure}
    \caption{\revisionCHI{An example of \textit{Coordination by Correcting} in \ParticipantFour{}'s human cooperation gameplay.}}
    \label{fig:gameplay-correction}
\end{figure}

\hc{}
\revisionCHI{When both the pilot and the copilot could act on the same \textit{game actions}, correction could be achieved by intervening on the shared \textit{game actions}.}
\revisionCHI{For example, \ParticipantFour{} often oversteered in one direction, and the copilot countersteered to adjust the trajectory.}
\revisionCHI{When the pilot and the copilot controlled different \textit{game actions}, the copilot could use one of the actions they controlled to correct the pilot's actions.}
\revisionCHI{For example, the copilot could
boost to resume movement when the user was not accelerating (\ParticipantFour{}), as shown in Fig.~\ref{fig:gameplay-correction}.}

\pa{}
The copilot occasionally corrected the pilot also during partial automation.
\revisionCHI{Typical corrective actions included boosting when the pilot did not accelerate forward (\ParticipantNine{}, \ParticipantTwelve{},\ParticipantSixteen{}),
and braking to reduce speed (\ParticipantFive{}). 
In particular, \ParticipantFive{} expressed appreciation for this braking assistance, as it helped control the vehicle:}
\revisionCHI{
\begin{displayquote}
\ParticipantFive{}: \it
[The software copilot] helps you really well. Even if you are accelerating, it brakes slightly and manages to keep your speed under control.
\end{displayquote}
}

\subsubsection{Ineffective Coordination}
\label{sssec:gameplay_interventions-cooperation_ineffective}

\hc{}
\revisionCHI{Players were sometimes unable to coordinate as a consequence of \quots{confusion due to unexpected actions} (Section~\ref{sssec:confusion-unexpected}).
In other cases, despite understanding each other's intentions, they were not able to coordinate the timing of their actions or the precision of their controls.}
For example, on one occasion, the copilot jumped too late with respect to \ParticipantFour{}'s movements, missing the goal.

\begin{figure}[htb]
    \centering
    \begin{subfigure}[t]{0.32\linewidth}
        \includegraphics[width=\linewidth]{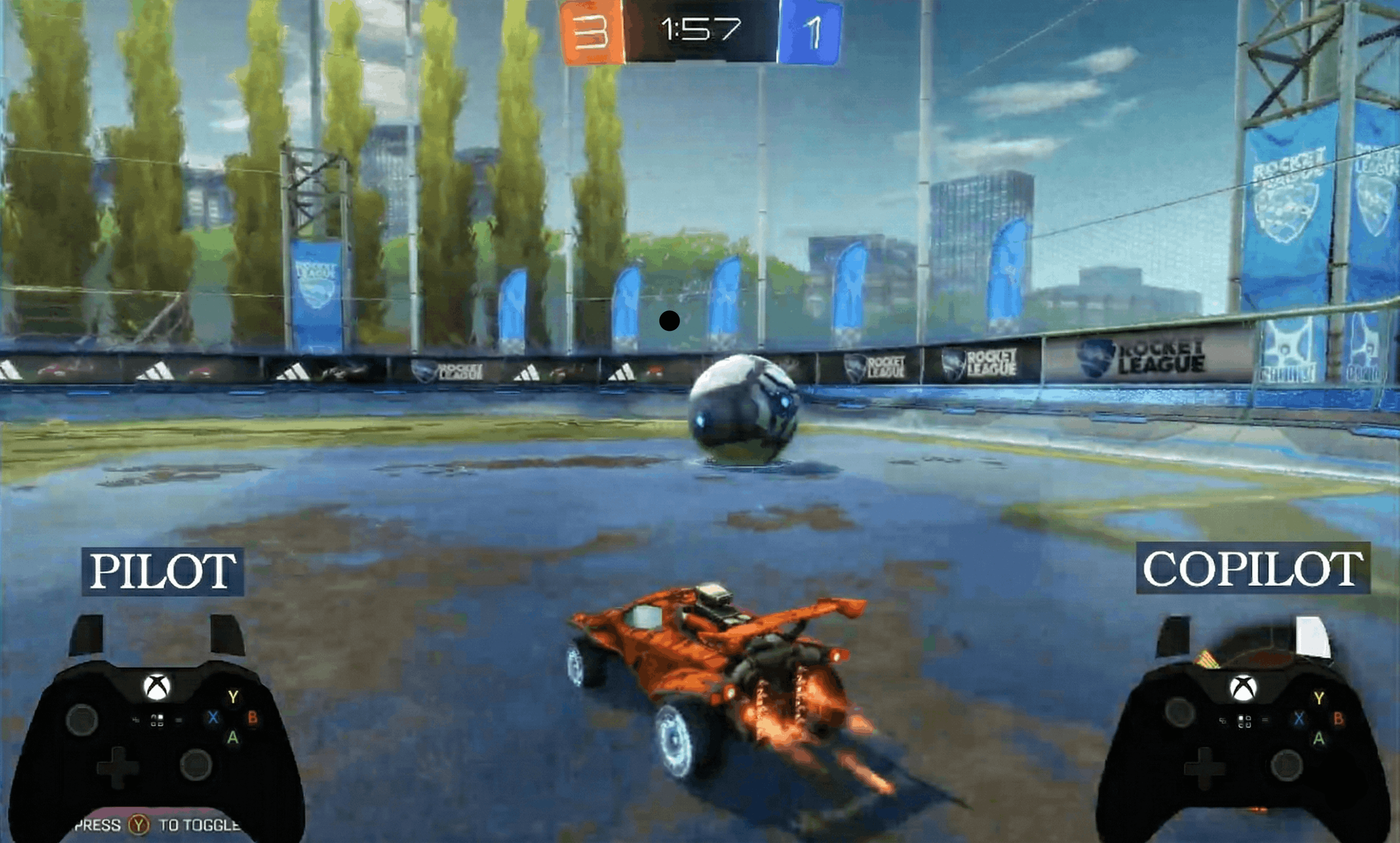}
        \caption{\revisionCHI{The copilot is heading towards the ball.}}
        \label{fig:gameplay-ineffective-coordination-pa-1}
        \Description{Screenshot of the augmented video used for the analysis of P12's partial automation session. P12 and their copilot are controlling an orange car. The car is moving forward, with the ball slightly to their right not far away.
        The pilot's controller shows no inputs.
        The copilot's controller shows the right trigger (acceleration) being pressed and the left trigger (steering) held slightly to the right.}
    \end{subfigure}
    \hfill
    \begin{subfigure}[t]{0.32\linewidth}
        \includegraphics[width=\linewidth]{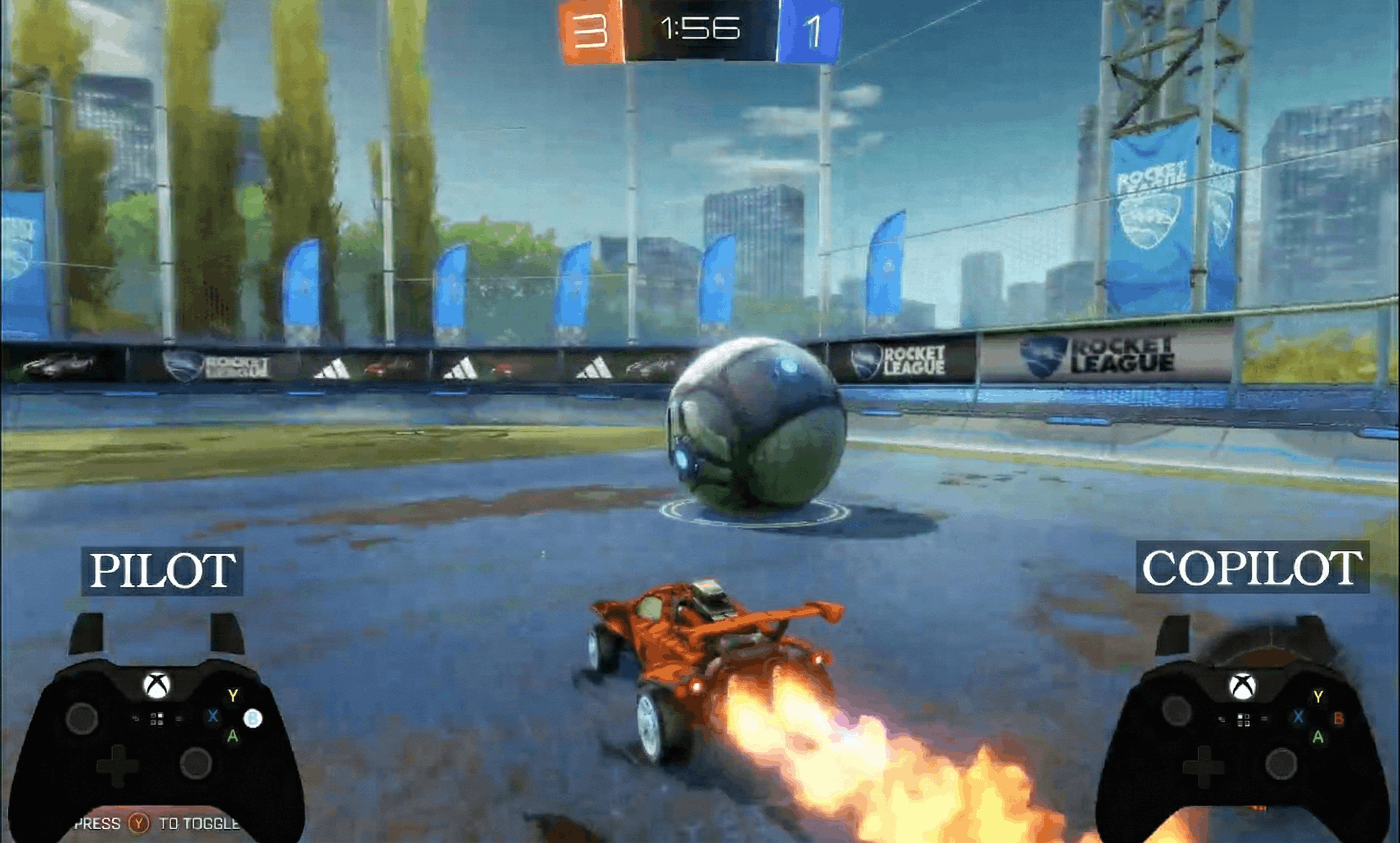}
        \caption{\revisionCHI{The pilot boosts.}}
        \label{fig:gameplay-ineffective-coordination-pa-2}
        \Description{Screenshot of the augmented video used for the analysis of P12's partial automation session. P12 and their copilot are controlling an orange car. The car is boosting forward, with the ball slightly to their right, close.
        The pilot's controller shows the B button (boost) being pressed.
        The copilot's controller shows the left trigger (steering) held slightly to the right.}
    \end{subfigure}
    \hfill
    \begin{subfigure}[t]{0.32\linewidth}
        \includegraphics[width=\linewidth]{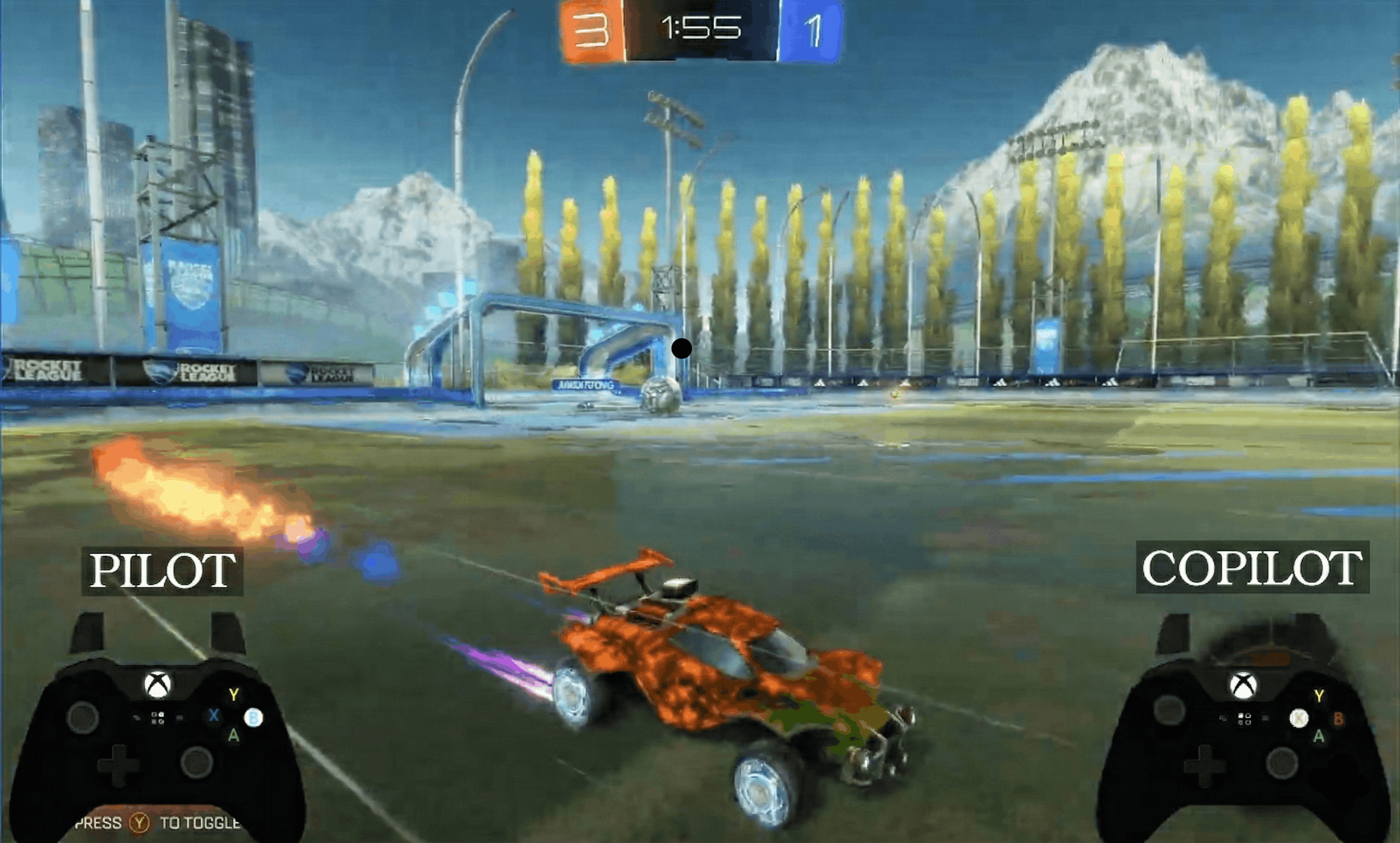}
        \caption{\revisionCHI{As the boost fuel diminishes, the copilot steers away from the ball towards a charging pad.}}
        \label{fig:gameplay-ineffective-coordination-pa-3}
        \Description{Screenshot of the augmented video used for the analysis of P12's partial automation session. P12 and their copilot are controlling an orange car. The car is boosting to the right, moving away from both the ball and the goal.
        The pilot's controller shows the B button (boost) being pressed.
        The copilot's controller shows the right trigger (acceleration) being pressed.}
    \end{subfigure}
    \caption{\revisionCHI{An example of \textit{Ineffective Coordination} in \ParticipantFifteen{}'s partial automation gameplay.}}
    \label{fig:gameplay-ineffective-coordination-pa}
\end{figure}

\pa{}
\revisionCHI{In partial automation, the software copilot’s use of boost and jump actions was sometimes unpredictable for the participants (Section~\ref{sssec:confusion-unexpected}), making coordination difficult.}
\revisionCHI{In other cases, ineffective coordination stemmed from unclear intent. For example, \ParticipantFifteen{} and \ParticipantSixteen{} delegated steering to the copilot but struggled to adapt to its behavior, which was not always clear to them, and over which they had no means of influence.
For example, \ParticipantFifteen{} could not understand why the copilot sometimes chose to head away from the ball (Fig.~\ref{fig:gameplay-ineffective-coordination-pa}).}

\subsubsection{Effect of the Software Copilot's Skills}
\label{sssec:gameplay_interventions-software-copilot-skills}
\revisionCHI{One notable difference between the support provided by the human copilot and the software copilot is that the software copilot was trained to emulate a highly skilled player who makes use of advanced strategies and can finely control its \textit{game actions} (Section~\ref{ssec:apparatus}).
In contrast, the human copilot was not an expert in the game.
For example, the software copilot was noted to steer with higher precision (\ParticipantFifteen{}, \ParticipantSixteen{}).}
\revisionCHI{Another difference is that the copilot had access to the complete game state, beyond what a human copilot could see on the video feedback. Therefore, the software copilot would execute jumps to hit the ball even when outside the camera view (\ParticipantSeven{}, \ParticipantNine{}, \ParticipantSixteen{}).}
Finally, the software copilot could react to changes in the game state (including the pilot's actions) $15$ times per second (Section~\ref{ssec:apparatus}).
This corresponds to a reaction time of $66.7ms$, which is lower than that of human players~\cite{dye2009increasing}.

For participants with previous experience with similar video games, adaptation to the copilot’s actions was generally quick and coordination remained mostly smooth (\ParticipantTwo{}, \ParticipantSix{}, \ParticipantEight{}).
Conversely, the software copilot's tactics sometimes resulted in unexplainable actions for a novice player, thus creating confusion (Section~\ref{ssec:confusion}) and hindering coordination (Section~\ref{sssec:gameplay_interventions-synergizing}).
For these reasons, participants expressed further interest in expanding the \systemFullName{}'s capabilities by offering personalization mechanisms to better adapt to individual playing styles and preferences (\ParticipantTwo{}, \ParticipantEleven{}).
\ParticipantFour{} envisioned a more sophisticated adaptive mechanism, drawing parallels to voice recognition systems that improve over time through exposure to individual users:
\begin{displayquote}
\revisionCHI{
\ParticipantFour{}: \it
It would be nice to have [a software copilot] that, as the user plays more and more the same video game, learns how to self-correct based on the errors that they usually make. Just like [the human copilot] did.
}
\end{displayquote}

\section{Discussion}
\label{sec:discussion}

Based on the results of our analysis, we discuss the effectiveness of partial automation (Section~\ref{ssec:discussion_effetiveness}), the dynamics of player collaboration (Section~\ref{ssec:discussion_collaboration}), and the design recommendations for a generalizable partial automation system (Section~\ref{ssec:discussion_design}).
Table~\ref{tab:findings} summarizes and enumerates the main findings.
Finally, we also discuss the limitations of our study (Section~\ref{ssec:limitations}).
\begin{table}[htb]
\footnotesize
%\centering
\caption{Main findings resulting from the study}
\label{tab:findings}
%\resizebox{\textwidth}{!}{%
\begin{tabular}{rll}
\toprule
\multicolumn{3}{l}{\hyperref[ssec:discussion_effetiveness]{\textbf{RQ1: Perception of Human Cooperation and Partial Automation}}} \\
\midrule
\textit{F1} & \textit{Accessibility} &
Human cooperation and partial automation are effective accessibility solutions \\
\textit{F2} & \textit{Autonomy} &
Partial automation does not require human assistance \\
\textit{F3} & \textit{Performance Gains} &
Shared control improves gameplay performance \\
\textit{F4} & \textit{Engagement} & 
Shared control maintains user engagement by preserving control over some \textit{game actions} \\
%\textit{F3} & \textit{Performance} &
%The software copilot can outperform a human copilot \\
%\textit{F4} & \textit{Interpretability} &
%The software copilot can adopt strategies that can be hard to interpret for novice players \\
\midrule
\multicolumn{3}{l}{\hyperref[ssec:discussion_collaboration]{\textbf{RQ2: Player Collaboration in Human Cooperation and Partial Automation}}} \\
\midrule
\textit{F5} & \textit{Synergy and Correction} &
Players collaborate by acting synergistically or by correcting one another \\
\textit{F6} & \textit{Ease of Coordination} &
The human copilot produces actions that are easier to interpret and align with \\
%\textit{F7} & \textit{Role of Dialogue} &
%Dialogue is central to human collaboration, and its absence is perceived as a limitation of \system{} \\
%absence is a significant limitation of partial automation \\
\midrule
\multicolumn{3}{l}{\hyperref[ssec:discussion_design]{\textbf{RQ3: Design Recommendations for Future Partial Automation Systems}}} \\
\midrule
\textit{F7} & \textit{Adaptation} &
The software copilot's play style should dynamically adapt to the pilot’s play style \\
\textit{F8} & \textit{Configuration} &
The software copilot's skill level should be manually configurable according to players preferences \\
\textit{F9} & \textit{Verbal Interaction} &
The system should support verbal communication \\
% for instructing, signaling, planning, and commanding \\
%\textit{F11} & \textit{Non-Verbal Interaction} &
%The system should support non-verbal interaction to substitute or reinforce verbal interaction \\
\textit{F10} & \textit{Interpretability} &
The system should provide real-time feedback what actions the copilot is taking and why \\
\bottomrule
\end{tabular}
%}
\Description{
TODO
}
\end{table}

\subsection{Perception of Human Cooperation and Partial Automation (\textbf{RQ1})}
\label{ssec:discussion_effetiveness}

Our study confirms previous findings that \textbf{human cooperation and partial automation are both perceived as effective solutions to make video games accessible [F1]} for people with disabilities~\cite{cimolino2021role} (Section~\ref{sssec:configuration-usefulness-shared}).
Indeed, all participants managed to play the game used in the study (Section~\ref{ssec:stimulus}), completing both gaming sessions.
This is particularly significant because many participants highlighted that they would not have been able to play the game without the provided support.
Additionally, \textbf{partial automation supports independent gameplay by removing the need for a human assistant [F2]}, unlike human cooperation~\cite{ahmetovic2026shared} (Section~\ref{sssec:ux-preferred_modality}).

Some participants stated that they would have been able to play the game without support.
However, they \textbf{recognized that playing in shared control was still effective [F3]} and reduced perceived cognitive load (Section~\ref{sssec:configuration-usefulness-shared}).
Partial automation was perceived as particularly effective, allowing the player to focus more on the actions they were controlling and improve their perceived gaming performance
%In fact, they perceived improvement in their performance 
%and a reduced workload 
(Section~\ref{sssec:ux-preferred_modality}).
%and more control of their actions (Section~\ref{sssec:configuration-usefulness-shared}).
%and \ParticipantSix{} also noted that they could focus more on the actions they were controlling (Section~\ref{sssec:configuration-usefulness-shared}).

Prior literature has observed that delegating gaming actions could lead to a lack of engagement~\cite{cimolino2021role,cimolino2023automation}.
In our study, we instead observed that \textbf{participants appeared engaged during gameplay [F4]}, in line with other works in the field~\cite{gonccalves2021exploring,loparev2014introducing}.
Some participants (\textit{e.g.}, \textbf{P5}) explicitly mentioned that this is because, despite some actions being delegated, they retained control over the actions they selected.
Other participants, instead, did not explicitly report this, but still showed a clear sense of engagement despite controlling only a few actions, such as \ParticipantSeven{}, who only controlled the steering, and \ParticipantFifteen{}, who only controlled jumps and boosts (Section~\ref{sssec:ux-sentiment}).

\subsection{Player Collaboration in Human Cooperation and Partial Automation (\textbf{RQ2})}
\label{ssec:discussion_collaboration}

Our analysis identifies two modalities through which the pilot and the copilot cooperate (Section~\ref{sssec:modes}) and analyzes the perceived ease of coordination (Section~\ref{sssec:in-game-interactions}).%, and the role of the dialogue during the collaboration (Section~\ref{sssec:dialogue}).

%\subsubsection{Collaboration Modalities}
\subsubsection{Collaborations Approaches: Synergy and Correction}
\label{sssec:modes}

In both human cooperation and partial automation, players adopt two modes of collaboration, rapidly switching between them depending on the game circumstances. %(Section~\ref{sssec:gameplay_interventions-synergizing} and Section~\ref{sssec:gameplay_interventions-correction}).
In the first mode, \textbf{both players act synergistically [F5]}: pilot and copilot contribute complementary actions toward the same goal (Section~\ref{sssec:gameplay_interventions-synergizing}).
For example, while the pilot steers toward the ball, the copilot may activate the boost to increase the vehicle’s speed and reach the ball faster.
In the second collaboration mode, \textbf{one player corrects the other’s actions [F5]}, compensating for inaccurate or incomplete actions and helping maintain gameplay effectiveness despite errors (Section~\ref{sssec:gameplay_interventions-correction}).
For example, if the pilot oversteers, the copilot may steer in the opposite direction to adjust the trajectory.
These two modes parallel a well-established distinction in human–human collaboration research.
Synergetic collaboration reflects the findings of Sacheli et al.~\cite{sacheli2018evidence}, who show that interacting individuals form a ``dyadic motor plan'', that unifies both players' expected actions.
This allows them to anticipate each other's outcomes and adapt their behavior to provide complementary inputs toward a common goal.
Corrective collaboration aligns instead with mechanisms of interpersonal action monitoring and compensatory behavior~\cite{sacheli2021mechanisms}, where one partner detects deviations from the shared goal and provides corrective inputs to maintain coordination.
To the best of our knowledge, this is the first paper that adapts these two concepts to the field of human-AI collaboration.

\subsubsection{Ease of Coordination With the Copilot}
\label{sssec:in-game-interactions}

%The first difference between human cooperation and partial automation concerns the different play styles adopted by the human and software copilots.
%Specifically, t
The software copilot was perceived by some participants to be more proficient at the game than the human copilot (Section~\ref{sssec:gameplay_interventions-software-copilot-skills}).
This also contributed to some participants feeling more in control and perceiving that they performed better (see [F3]) (Section~\ref{sssec:ux-preferred_modality}).
However, other participants reported difficulties in interpreting the rationale behind the software copilot's behavior.
Consequently, they preferred human cooperation because the \textbf{\textit{game actions} produced by the human copilot were easier to interpret and, ultimately, easier to align with [F6]} (Section~\ref{sssec:confusion-unexpected}).
This difference can be explained by the fact that the human copilot was not an expert player of \textit{Rocket League} and adopted a gaming style similar to the participants, also dynamically adapting their style of play to the pilot (Section~\ref{sssec:gameplay_interventions-software-copilot-skills}).
Such adaptation was not possible in partial automation (see [F7] below), as the model was pre-trained to use advanced tactics, for example, to jump while boosting to better propel forward, to which participants with limited game expertise were not able to respond effectively (Section~\ref{ssec:confusion}).
Additionally, the software copilot's behavior could not be adjusted dynamically based on the pilot's skills (Section~\ref{ssec:stimulus}).

\subsection{Design Recommendations for Future Partial Automation Systems (\textbf{RQ3})}
\label{ssec:discussion_design}

Our study highlights three requirements that future partial automation systems should address to improve support for players with upper-limb impairments.

\subsubsection{Adaptation to the Pilot’s Play Style}    
\textbf{The software copilot should adapt to the pilot’s play style [F7]}, either automatically or based on explicit guidance from the pilot (Section~\ref{sssec:gameplay_interventions-software-copilot-skills}).
Here, ``play style'' refers to the way the copilot plays, such as aggressive, defensive, or timing of specific actions, aligning with the player’s personal approach.
Two possible approaches are: providing multiple pre-trained game agents to implement different play styles so that the player can explicitly select one, or designing a single agent capable of dynamically adapting to the player’s style.

%TC:ignore
\begin{comment}
\revisionCHI{On this note, our study uncovers that \textbf{partial automation systems should dynamically adapt to the player}, a feature that is currently missing in \system{}, which cannot align its game style with that of the pilot (Section~\ref{sssec:gameplay_interventions-software-copilot-skills}), nor allow the pilot to indicate how they wish to be supported (Section~\ref{sssec:communication-planning})}.
More in general, partial automation in the \systemFullName{} does not provide communication facilities to support the understanding between the pilot and the copilot.
Such support would also be needed to allow the system to communicate the intent to perform actions (Section~\ref{sssec:communication-signaling}) or to strategize with the player (Section~\ref{sssec:communication-planning}).
\end{comment}
%TC:endignore

\subsubsection{Configurable Level of Support}
\textbf{The participants expressed the need to calibrate the \textit{game agent}’s skill level [F8]} (Section~\ref{sssec:gameplay_interventions-software-copilot-skills}).
Unlike play style, ``skill level'' refers to how competent or effective the copilot is relative to the player.
If the agent is too skilled compared to the player, its actions may be difficult to interpret; conversely, some participants appreciated a highly skilled agent because it gave them a stronger sense of competence (Section~\ref{sssec:ux-preferred_modality}).
The level of support could therefore be selected directly by the participant or automatically inferred by the agent.
Automatically adapting the level of support to the pilot’s abilities was also mentioned in prior literature as a way to increase the acceptability of partial automation in multiplayer games~\cite{ahmetovic2026shared}.

%\subsubsection{Verbal Interaction}
%\textbf{Participants emphasized the need for partial automation systems to incorporate verbal communication [F10]} features and engage in dialogue, in particular about instructing, signaling, planning, and commanding (Section~\ref{ssec:communication}).

\subsubsection{Verbal Interaction}
The lack of verbal interaction contributed to participants perceiving the partial automation as less natural (Section~\ref{sssec:ux-preferred_modality}).
For this reason, \textbf{partial automation systems should be able to incorporate verbal communication [F9]}, for example, to allow the player to verbally select a play style or to tune the skill level during the game (Section~\ref{sssec:gameplay_interventions-software-copilot-skills}).

%\subsubsection{Non-Verbal Interaction}
%\textbf{The copilot should also allow non-verbal interaction [F11]} either as a substitute for verbal dialogue or as reinforcement. For example, a verbal message signaling a game element to the player could be paired with an on-screen indicator (\textit{e.g.}, a red box around the object) (Section~\ref{sssec:communication-signaling}).
	
\subsubsection{Interpretability of the Copilot's Behavior}
We observed episodes of collaboration confusion when players misattributed actions to themselves or to the copilot (Section~\ref{ssec:confusion}).
To address this issue, \textbf{partial automation systems should provide clear, real-time feedback about which actions are being performed by the copilot and why [F10]}, thereby reducing uncertainty and enhancing trust. Participants suggest achieving this with on-screen indicators or verbal feedback (Section~\ref{sssec:confusion-unexpected}).

\subsection{Limitations}
\label{ssec:limitations}

We discuss six main limitations: those related to the design and implementation of the \systemFullName{} (Section~\ref{sssec:limitations-framework}), to participant recruitment (Section~\ref{sssec:limitations-participants}), to the game stimulus used (Section~\ref{sssec:limitations-game}), to the selection of the human copilot (Section~\ref{sssec:limitations-copilot}), to the choice of configuration by participants (Section~\ref{sssec:limitations-configuration}), and those related to the scope of our analysis (Section~\ref{sssec:limitations-analysis}).

\subsubsection{Framework Design and Implementation}
\label{sssec:limitations-framework}

The design and implementation of \system{} focused on enabling gameplay interactions between a human pilot and a human or software copilot, and on being able to support arbitrary third-party video games (Section~\ref{sec:system}).
%One system limitation, which was strongly perceived by the participants, is that the \textit{game agent} employed for partial automation only generates \textit{game actions} and is not capable of communicating with the pilot (Section~\ref{ssec:communication}).
%In particular, in the absence of a feedback system to communicate the intent of the copilot's actions, the actions of the software copilot can be confusing (Section~\ref{sssec:communication-signaling}).
One system limitation is that the level of support provided by the software copilot cannot be tuned to the pilot's needs and preferences or adapted to the pilot's play style.
%Furthermore, the actions of the software copilot can be confusing in the absence of a feedback system to communicate the intent of the copilot's actions (Section~\ref{sssec:communication-signaling}).
This was particularly relevant because the \textit{game agent} was trained with the aim of maximizing its own ability to compete with other bots, and action interpretability by other players was not a training objective (Section~\ref{sssec:gameplay_interventions-software-copilot-skills}).
Furthermore, the actions of the software copilot can be confusing in the absence of a feedback system to communicate the intent of the copilot's actions (Section~\ref{ssec:confusion}).

\subsubsection{Study Participants}
\label{sssec:limitations-participants}

Our study involved $13$ representative participants, a number in line with the current accessibility literature~\cite{mack2021we} (Section~\ref{ssec:participants}).
The participants had various mobility impairments and were distributed across genders, age ranges, prior gaming experience, and familiarity with assistive technologies.
%\revisionCHI{While this heterogeneity introduces potential confounding factors that may limit the validity of our results, the sample also shows an imbalance in impairment types, with $7$ of the $13$ participants having muscular dystrophy, which may limit the generalizability of our findings to individuals with other types of mobility impairments.}
Although this heterogeneity helps to support the generalizability of our findings, it can also introduce potential confounding factors.
Another limitation is that $7$ of the $13$ participants have muscular dystrophy, which can limit the generalizability of our findings.
Additionally, only a few participants had prior experience with the game, which allowed us to investigate how novices benefit from the shared control support, but not the effect on expert players.
Another limitation is that all participants were recruited in the same geographical region, thus may not be culturally representative.
\revisionCHI{These limitations suggest that further research with larger samples would be necessary to improve the generalizability of our findings.}

\subsubsection{Study Duration}
\label{sssec:limitations-game}

The game we selected as a stimulus for the experiments was simple in terms of overall objectives and main controls, which allowed us to conduct the experiments without prolonged training (Section~\ref{ssec:apparatus}).
However, the limited time of the experiment in some cases was not sufficient to form a complete understanding of the game.
Indeed, more advanced game mechanics or tactics, including those sometimes employed by the software copilot, were not immediately clear to the participants (Section~\ref{sssec:gameplay_interventions-software-copilot-skills}).
%, and, indeed, a strong learning effect was observed between the two game sessions.
\revisionCHI{For this reason, a follow-up longitudinal study could examine how performance changes as participants gain familiarity with the game and the cooperative system.
In particular, such longitudinal studies would be needed for future experiments involving more complex video games.
}

\subsubsection{Single Copilot}
\label{sssec:limitations-copilot}

We conducted our experiments with representative participants as pilots, but only one human copilot, who was also the experiment supervisor (Section~\ref{ssec:participants}).
By definition, therefore, we were not able to assess how different levels of copilot support would impact the collaboration in human cooperation.
There was also a potential copilot's learning effect, with respect to the game, and with respect to providing support in human cooperation, that perhaps resulted in the copilot providing diverse levels and quality of support between the first and the last participants.
Additionally, none of the participants were familiar with the human copilot, whereas those who play in human cooperation usually have familiarity with the copilot, a factor that impacts their mutual ability to play together~\cite{ahmetovic2021replay}.

\subsubsection{Impact of Configuration Choice}
\label{sssec:limitations-configuration}

\revisionCHI{In our study, participants were free to choose the configuration they preferred to interact with the game.
This allowed them to adapt the controls to their needs and preferences.
As discussed in Section~\ref{ssec:configuration}, prior familiarity with traditional gamepads or with \textit{Rocket League} influenced these choices, with more experienced players typically retaining control over a higher number of \textit{game actions}.
These configuration choices may also have influenced the study's outcomes.
For example, participants who controlled fewer \textit{game actions} (\textit{e.g.}, \ParticipantSeven{}, \ParticipantNine{}, \ParticipantTen{}, \ParticipantSixteen{}) relied more heavily on the copilot and had less direct control over gameplay, which may have affected their sense of agency. 
Conversely, participants who opted to control more \textit{game actions} (\textit{e.g.}, \ParticipantSix{}, \ParticipantEight{}) may have benefited less from the copilot's assistance, potentially increasing cognitive and motor demands.
Participants generally reported being satisfied with their chosen \textit{game actions} subdivision.
However, further investigation is needed to assess how different configurations influence players’ subjective experience.}

\subsubsection{Scope of the Analysis}
\label{sssec:limitations-analysis}

In this paper, we focused exclusively on understanding how \textit{assistance by playing}~\cite{ahmetovic2026shared} is provided, that is how a copilot can support a pilot by intervening on the game controls.
Under this lens, we explored how participants played under shared control, analyzing their actions, coordination patterns, and gameplay strategies.
Other relevant aspects of human cooperation, such as those related to the communication between the pilot and the copilot, were not addressed.
While these aspects are outside the scope of the present work, their understanding can be relevant to form a complete model of shared control for game accessibility and will be investigated as future work.

\section{Conclusions and Future Work}
\label{sec:conclusions}
Our study provides evidence that both human cooperation and partial automation enable players with upper-limb impairments to access video games.
The analysis shows that human cooperation fosters rich verbal communication and strategy negotiation, often generating complicity but also dependence on the human copilot.
Partial automation, on the other hand, increases autonomy and, in some cases, enhances the sense of competence, though it occasionally introduces collaboration confusion when actions are misattributed. 
These results highlight that human cooperation and partial automation, as implemented in \system{}, support accessibility through distinct collaboration dynamics, offering complementary benefits and limitations.
The advantages of human cooperation, particularly its natural communication and adaptability to the user's style of play, can inform the design of partial automation systems that collaborate more naturally with the player.
%This paper also presents the \systemFullName{}, which is highly configurable to accommodate both forms of shared control, as well as hybrid configurations, and can therefore be employed as a probe to explore new interaction paradigms. Its public availability further increases its potential impact, enabling replication of our study and fostering future research on game accessibility and shared control systems.
%
This paper also presents the \systemFullName{}, which provides a configurable experimental probe for studying these shared-control dynamics across human and software copilots.

We identify four main directions for future work.
First, building on the methodology presented in this study, future research could investigate shared control as an accessibility strategy in different contexts, for example, for people with disabilities other than upper-limb impairments.
Another line is to explore accessibility to different game genres, including slower-paced titles compared to \textit{Rocket League}, which may elicit distinct collaboration dynamics. 
Moreover, future studies could involve pairs of participants who already play as pilot and copilot, to better investigate collaboration within consolidated dyads.
Longitudinal studies could also provide insights into how players’ perceptions and strategies evolve over time when using human cooperation and partial automation.

Second, future research could analyze the forms of dialogue between the pilot and the copilot.
This could inform the design of a software copilot that, in addition to providing \textit{assistance by playing}, can also assist by verbally interacting with the player, for example, to signal elements of interest in the game or to suggest how to resolve particularly difficult situations~\cite{ahmetovic2026shared}.

A third research direction involves the design of advanced software copilots that better emulate the human copilot.
\revisionCHI{This includes modeling the player’s abilities and preferred strategies so that the copilot can tailor its interventions and avoid adopting gameplay strategies that exceed the player’s skill level.
To achieve this, explicit communication mechanisms could be introduced to negotiate game strategy and provide support beyond direct in-game actions.
Additionally, improving transparency would help the pilot more easily interpret and anticipate the copilot’s decisions.
Another approach could integrate basic coaching behaviors, reflecting the supportive role that the human copilot naturally assumed in our study.}

Finally, our findings can be extended to domains outside of entertainment.
Potential applications include rehabilitation, where a software copilot could support therapeutic exercises, and educational environments, where shared control could scaffold learning activities.

\bibliographystyle{ACM-Reference-Format}
\bibliography{main}

%TC:ignore
\appendix

\section{\systemFullNameTitle{} Implementation}
\label{appendix:implementation}

\revisionCHI{We discuss the main implementation details of the \systemFullName{}.
In particular, we explain how we transmitted \textit{input commands} to the game (Appendix~\ref{appendix:virtual-controller}), how we masked users' inputs (Appendix~\ref{appendix:input-masking}), the two merging policies used in our study (Appendix~\ref{appendix:policies}), and the \textit{Rocket League} game adaptation (Appendix~\ref{appendix:game-adaptation}).}

\subsection{\revisionCHI{Virtual Controller and Input Simulation}}
\label{appendix:virtual-controller}

\revisionCHI{To execute the \textit{input commands} generated by the \textit{command arbitrator}, the \systemFullName{} employs a virtual device recognized by the game as a standard physical controller.
We used the \textit{vgamepad}\footnote{\revisionCHI{\url{https://github.com/yannbouteiller/vgamepad}}} Python library, which interfaces with the \textit{ViGEmBus} (Virtual Gamepad Emulation Bus) driver\footnote{\revisionCHI{\url{https://github.com/nefarius/ViGEmBus}}}.
This setup enables the emulation of a standard \textit{Xbox 360} controller, ensuring native compatibility with any video game supporting gamepad input.}

\revisionCHI{Each \textit{input command} generated by the \textit{command arbitrator} consists of a pair of \texttt{input\_command} and \texttt{intensity} values.
The \texttt{input\_command} value identifies the controller input (\textit{i.e.}, button, trigger, or analog stick axis) that should perform the input.
The \texttt{intensity} value defines the magnitude with which the input is activated.}
\revisionCHI{The interpretation of \texttt{intensity} varies based on the type of the associated \texttt{input\_command}:}
\begin{itemize}
    \item \revisionCHI{\textbf{Buttons}: when \texttt{intensity} is $1$ the button is pressed, else it is released.}
    \item \revisionCHI{\textbf{Triggers}: the continuous value of \texttt{intensity} in $[0, 1]$ is mapped to the floating-point format required by the \textit{vgamepad} library.}
    \item \revisionCHI{\textbf{Analog Sticks}: \texttt{intensity} is a pair of values in $[-1, 1]$ representing the stick movement on both axes (\textit{e.g.}, $<-1, 0>$ represents the stick full on the left).}
\end{itemize}

\subsection{\revisionCHI{Physical Controller Input Masking}}
\label{appendix:input-masking}

\revisionCHI{To ensure that the game processes \textit{input commands} exclusively from the \textit{virtual controller}, we implement an input masking mechanism for the player's physical device.
Without this step, the game would receive input from both the physical and the virtual controllers.}

\revisionCHI{We utilize \textit{HidHide}\footnote{\revisionCHI{\url{https://github.com/nefarius/HidHide}}}, an open-source utility that acts as a filter driver for Human Interface Devices (HID).
This tool allows us to ``hide'' the physical controller from the operating system and the game, while still allowing the \system{} to read its raw inputs via a whitelist mechanism.}

\revisionCHI{Currently, \textit{HidHide} operates as a third-party dependency that must be activated and configured separately from the main framework.
However, given its open-source nature, it may be natively integrated into future versions of the \systemFullName{} to provide a unified solution.}

\subsection{\revisionCHI{Adopted Merging Policies}}
\label{appendix:policies}

\revisionCHI{The \systemFullName{} creates an abstraction layer where inputs from different sources (human players and \textit{game agents}) are normalized into \textit{input entries}.
Each entry $e$ consists of a value $v \in \mathbb{R}$, a confidence level $c \in [0, 1]$, and a role $r \in \{\text{Pilot}, \text{Copilot}\}$.
To determine the final value sent to the virtual controller, the \textit{command arbitrator} applies a configurable \textit{merging policy} for each \textit{game action}.
In our study we adopted two different \textit{merging policies}.}

\subsubsection{\revisionCHI{Binary \textit{Game Actions} Merging Policy}}

\revisionCHI{For jumping, boosting, and handbraking \textit{game actions} in \textit{Rocket League}, we adopted a policy implementing an inclusive disjunction: the action is triggered if at least one actor provides a non-zero input.
Formally, given a set of input entries $E$, the output value $v_{out}$ is defined as:
\begin{equation}
    v_{out} = 
    \begin{cases} 
        1 & \text{if } \exists e \in E : e.v \neq 0 \\
        0 & \text{otherwise}
    \end{cases}
\end{equation}
This ensures that if either the human player or the software agent executes an action, the command is registered by the game.}

\subsubsection{\revisionCHI{Continuous \textit{Game Actions} Merging Policy}}

\revisionCHI{For steering, acceleration, and braking \textit{game actions}, we adopted an \quots{unsupervised} mechanism~\cite{cimolino2022two} in which no actor is given inherent priority.
Thus, the merging policy outputs the average of the input values.}

\subsection{\revisionCHI{Rocket League Game Adaptation}}
\label{appendix:game-adaptation}

\revisionCHI{In order to compare partial automation to human cooperation in our study, we developed a specific game adaptation for the \textit{Rocket League} video game~\cite{rocketLeague}.
The adaptation consists in two distinct components: the \textit{game state reader} (Appendix~\ref{appendix:game-state-reader}), responsible for extracting the state of the game, and the \textit{game agent} (Appendix~\ref{appendix:game-agent}), which generates new \textit{game actions} based on the extracted \textit{game state}.}

\subsubsection{\revisionCHI{Game State Reader}}
\label{appendix:game-state-reader}

\revisionCHI{
We developed a custom C/C++ plugin using the \textit{BakkesMod SDK}~\cite{bakkesMod}, which allows direct access to \textit{Rocket League}'s internal memory data structures.
Based on the information needed for the \textit{game agents}' decision-making process, we selected and extracted only the relevant elements of the game state.
The plugin hooks into the main rendering loop, gathers the selected data every frame (the game runs at $120$ FPS), serializes it to JSON, and transmits it to the \textit{game agents} via a local socket connection.}

\revisionCHI{The extracted state is composed of five primary fields: the ball state, the cars' states (one for each player in the match), team information, general game information, and boost pad states.}

\paragraph{\revisionCHI{Ball State}}
\revisionCHI{The ball state captures the complete physical representation of the ball and contains:}
\revisionCHI{\begin{itemize}
    \item Location: 3D position of the ball in the game world coordinate system (X, Y, Z in Unreal Engine\footnote{\revisionCHI{The game engine used to develop \textit{Rocket League}.}} units).
    \item Rotation: orientation of the ball expressed as pitch, yaw, and roll angles in radians, normalized to the range $[-\pi, \pi]$.
    \item Velocity: linear velocity vector representing the ball's speed and direction.
    \item Angular velocity: rotational velocity of the ball around each axis.
\end{itemize}}

\paragraph{\revisionCHI{Car States}}
\revisionCHI{For each player's car, the game state specifies:}
\revisionCHI{\begin{itemize}
    \item Physics data: complete physical state of the car (location, rotation, velocity, angular velocity), identical in structure to the ball state.
    \item Team ID: team identifier (0 for blue team, 1 for orange team).
    \item Demolished: boolean value indicating whether the car has been destroyed and is awaiting respawn.
    \item Ground contact: boolean value indicating ground contact.
    \item Jumped: boolean value indicating if the car has executed a jump.
    \item Boost: remaining boost fuel percentage.
    \item Is bot: distinguishes AI-controlled opponents from human players.
\end{itemize}}

\paragraph{\revisionCHI{Team Information}}
\revisionCHI{Each team's data contains:}
\revisionCHI{\begin{itemize}
    \item Team ID: team identifier (0 for blue team, 1 for orange team).
    \item Score: current goal count for the team.
\end{itemize}}

\paragraph{\revisionCHI{Game Information}}
\revisionCHI{The general match state only includes the seconds elapsed since the match start.}

\revisionCHI{Additionally, the game state includes a field that indicates the current UI state (in-game, paused, other), allowing the system to pause input processing when the game is not actively being played.}

\paragraph{\revisionCHI{Boost Pad States}}
\revisionCHI{Finally, the game state tracks the availability of each boost pad for collection.
Boost pads are items placed in fixed locations across the arena that replenish players' boost fuel when stepped on.
After being collected, boost pads are deactivated for a specified period.}

\subsubsection{\revisionCHI{Game Agent Integration}}
\label{appendix:game-agent}

\revisionCHI{To implement the partial automation \textit{game agent}, we integrated \textit{Nexto}\footnote{\revisionCHI{\url{https://github.com/Rolv-Arild/Necto}}}, a state-of-the-art reinforcement learning bot for \textit{Rocket League}.
\textit{Nexto} utilizes a neural network trained to play the game at a superhuman level, capable of executing complex mechanics and strategic positioning.}

\revisionCHI{The integration follows a two-step process.}

\paragraph{\revisionCHI{Inference Loop}}
\revisionCHI{Although the game state is extracted at $120$ FPS, the \textit{game agent} operates at a lower frequency to match its training conditions.
We transmit the state to the agent every 15 frames.
Upon receiving the \textit{game state}, the agent performs a forward pass of its neural network to predict the optimal action vector $\mathbf{a}_{bot} \in \mathbb{R}^8$, which includes continuous values for throttle, steer, pitch, yaw, and roll, and boolean probabilities for jumping, boosting, and handbraking.
To ensure smooth control during the intervals between updates (the 14 intermediate frames), the last computed action vector is repeated.}

\paragraph{\revisionCHI{Action Masking}}
\revisionCHI{The native \textit{Nexto} bot is designed to control the car autonomously.
To use it as a \textit{software copilot}, we implemented an output masking layer before the \textit{command arbitrator} that filters out all game actions except those assigned to the software copilot in the configuration.}

\section{Experimental Methodology}
\label{appendix:protocol}

\revisionCHI{This section presents additional details on the experimental methodology adopted in our study.}

\subsection{Design}

The study was conducted in person at \anon[our laboratory]{the EveryWare laboratory} or at a location convenient for the participant (\textit{e.g.}, an association's headquarters).
Date, time, and location were agreed upon with each participant.
In addition to the participant, at least one supervisor who is a member of the research group was always present during the study.
At any point during the study, the supervisor was required to answer any additional questions posed by the participant, if able; otherwise, they had to ask the participant to contact the research leader.

The study was be repeated with the same structure for each participant. It was conducted in 5 phases:

\paragraph{Phase 1 - Informed Consent}

Before the study, the candidate participant who meets the inclusion criteria (Appendix~\ref{appendix:recruitment}) was informed of the research objectives and the methods by which the study was to be conducted.
If the candidate participant consented to participate, they were provided with the informed consent and the privacy information sheet.
The supervisor ensured that the participant read and signed both documents.
The signatures for the informed consent and data processing authorization were collected on paper.
In cases where participants were unable to sign the informed consent, the participant's authorization was recorded in audio format.

\paragraph{Phase 2 - Demographic Questions}

The supervisor orally administered initial demographic questions to the participant.
These questions aimed to collect demographic data about the participant and, in the subsequent analysis phase, allowed for the study of how such data influenced the experimental results.

\paragraph{Phase 3 - Introduction to the Game and System Configuration}

The supervisor explained to the participant how the \textit{Rocket League} video game works and which \textit{game actions} it supports.
The explanation took place through a pre-recorded video.
Then, the participant was asked which \textit{game actions} they wished to control and how.
To control \textit{game actions}, the participant had access to:
\begin{itemize}
    \item A standard \textit{Xbox} controller~\cite{xboxController}.
    \item An \textit{Xbox Adaptive Controller}~\cite{xboxAdaptiveController} along with additional supporting peripherals:
    \begin{itemize}
        \item A \textit{Logitech Adaptive Gaming Kit}~\cite{logitechAdaptiveGamingKit}
        \item A \textit{Logitech Joystick Extreme 3D Pro}~\cite{logitechJoystick}
    \end{itemize}
\end{itemize}
The participant could freely choose how to control each \textit{game action}, which ones to leave to complete system control, and which ones to share control of instead.
Once the \textit{game actions} were chosen, the supervisor configured the necessary peripherals and \system{} accordingly.

The participant could try the chosen configuration in human cooperation.
During the trial, the second controller was used by the supervisor.
At the end of the trial, the participant could change the chosen configuration.
Once confirmed, it could not be changed for the entire duration of the study.

Before ending this phase, the supervisor started the audio and video recording of the experiment.

\paragraph{Phase 4 - Gaming Through Shared Control}
    
Two gaming sessions were conducted:
\revisionCHI{\begin{itemize}
    \item Human cooperation session, where the second controller was controlled by the supervisor.
    \item Partial automation session, where the second controller was controlled by a software agent.
\end{itemize}}
The order of the sessions was based on a Latin Square design.

Each session lasted between $5$ and $10$ minutes.

At the end of each session, two questionnaires were administered to the participant:
\begin{itemize}
    \item NASA Task Load Index (NASA-TLX)\cite{hart2006nasa}, to evaluate the participant's fatigue during the game.
    \item Player Experience of Needs Satisfaction (PENS)\cite{ryan2006motivational}, to evaluate the user experience during the game.
\end{itemize}
Both questionnaires are available and validated for the English language.
While a validated version of NASA-TLX in \anon[the language spoken by the research team]{Italian} is available\anon[]{~\cite{bracco2008metodi}}, the PENS questionnaire was manually translated by the research team for \anon[participants speaking the same language as the researchers]{Italian participants}.
Both questionnaires were provided in paper format.
For transparency, we note that these questionnaires were administered during the study, but are not analyzed in this paper, because no statistically significant differences emerged, and the focus here is on the qualitative aspects of how people with disabilities play in human cooperation and partial automation.

\paragraph{Phase 5 - Interview on Gaming Experience}

At the conclusion of the study, a semi-structured interview was conducted to explore the participants' gaming experiences with the two shared control modalities.
The interview was recorded, and the responses were subsequently transcribed.

During the interview, an initial set of questions was always posed, with the possibility of further probing based on participants’ responses.
The initial set of questions was informed by previous literature on shared control assistance~\cite{cimolino2022two, cimolino2023automation, gonccalves2021exploring}, and designed to address the study’s research questions.
The list of pre-defined questions is provided below.
For each, we also indicate the research question it addresses.
\begin{itemize}
   \item[RQ1] Which of the two gaming modalities did you prefer? Why?
   \item[RQ1] Was the game actions subdivision effective? If you were to continue playing, would you change it?
   \item[RQ1] During the AI-supported game, did you feel in control of the game actions? Why?
   \item[RQ1] Did you have difficulties or moments of frustration during the game? If so, which ones and in what context?
   \item[RQ1] If you had this system available at home, would you continue using it?
   \item[RQ1] Would you be able to play the game without the support of the artificial intelligence?
   \item[RQ2] Were the actions performed by the artificial intelligence always clear and predictable to you?
   \item[RQ3] What would you change or improve in the system to make it more suitable for your needs?
   \item[RQ3] How do you think the system could be modified to improve the [modality the participant liked the least]?
\end{itemize}

Each test had a maximum duration of approximately one hour.
Participants could discontinue participation in the study at any time, without having to provide any explanation.
Table~\ref{tab:timing} reports an estimate of the time required for each phase of the study.
\begin{table}[htb]
\centering
\caption{\revisionCHI{Estimated duration of the study.}}
\label{tab:timing}
\begin{tabular}{lc}
\hline
\textbf{\revisionCHI{Phase}} & \textbf{\revisionCHI{Estimate (minutes)}} \\ \hline
\revisionCHI{Demographic questionnaire} & \revisionCHI{10} \\
\revisionCHI{Game configuration}        & \revisionCHI{15} \\
\revisionCHI{First gaming session}      & \revisionCHI{10} \\
\revisionCHI{First questionnaire}       & \revisionCHI{5}  \\
\revisionCHI{Second gaming session}     & \revisionCHI{10} \\
\revisionCHI{Second questionnaire}      & \revisionCHI{5}  \\
\revisionCHI{Final interview}           & \revisionCHI{10} \\ \hline
\textbf{\revisionCHI{Total}}            & \revisionCHI{60} \\ \hline
\end{tabular}
\Description{
A table reporting the estimated duration of each phase of the study. Each row corresponds to one phase. The columns express the name of the phase and an estimated time duration, in minutes. The last row reports the sum of the other rows.
}
\end{table}

\subsection{Participants' Recruitment}
\label{appendix:recruitment}

Participants were recruited through convenience sampling in three ways:
\begin{itemize}
    \item Through third-party associations already in contact with people with disabilities. In this case, we asked the association to handle participant recruitment, thus avoiding direct contact with them until the time of the study.
    
    \item By contacting participants from one of our previous studies on similar topics. This method was only used with participants who had expressed a favorable opinion about being contacted again.
    
    \item Through snowball sampling, we asked study participants to share our contact information with their acquaintances, so that they could contact us if they were interested in participating in the study.
\end{itemize}
Participation in the study was on a voluntary basis.

\subsubsection{Recruitment Criteria}

To access the study, participants had to be people with an upper-limb disability that causes them difficulty in using a video game controller.
An exclusion criterion was the presence of multiple disabilities, as they could be a confounding factor during data analysis.
No previous experience with shared control technologies was required.

All participants had to speak \anon[languages spoken by the researchers]{Italian or English}.

Minor participants were admitted to the study with the consent of their parents or legal guardians.
In any case, participants had to be at least $13$ years old.
No maximum age was specified for participating in the study.

\revisionCHI{Since each participant in the study played \textit{Rocket League} in both human cooperation and partial automation, any previous experience with the game was not considered an exclusion criterion, as this should have influenced both gaming sessions equally.}

%TC:endignore
\end{document}